\documentclass[12pt, draftclsnofoot, onecolumn]{IEEEtran}
\ifCLASSINFOpdf
\else
\fi
\usepackage[utf8]{inputenc}
\usepackage{amsmath,amssymb,amsfonts}
\usepackage{cite}

\usepackage{algorithmic}
\usepackage{stfloats}
\usepackage{graphicx}
\usepackage{textcomp}
\usepackage{xcolor}
\usepackage{pifont}
\usepackage{amsthm}
\newtheorem{theorem}{Theorem}

\usepackage{hyperref}
\usepackage{mathrsfs}
\usepackage{booktabs}
\usepackage{hyperref}
\usepackage{cases}
\usepackage{comment}
\usepackage{empheq} 
\usepackage{caption}
\usepackage{subcaption}

\allowdisplaybreaks

\usepackage{url}  
\usepackage{graphicx}  
\usepackage[ruled,vlined,linesnumbered]{algorithm2e}
\usepackage{setspace}
\usepackage{floatpag} 
\usepackage{float}
\floatpagestyle{empty}

\usepackage{listings}

\begin{document}
\captionsetup[figure]{labelfont={rm},labelformat={default},labelsep=period,name={Fig.}}
\title{Reconfigurable Intelligent Surfaces Aided mmWave NOMA: Joint Power Allocation, Phase Shifts, and Hybrid Beamforming Optimization}
%
%
%

\author{Yue Xiu,~Jun Zhao,~\IEEEmembership{Member,~IEEE},
Wei Sun,~\IEEEmembership{Student Member,~IEEE},\\ ~Marco~Di~Renzo,~\IEEEmembership{Fellow,~IEEE},
~Guan Gui,~\IEEEmembership{Senior Member,~IEEE},\\
~Zhongpei Zhang,~\IEEEmembership{Member,~IEEE},~Ning Wei,~\IEEEmembership{Member,~IEEE}\\
\thanks{Yue Xiu, Zhongpei Zhang, and Ning Wei are with National Key Laboratory of Science and Technology on Communications, University of Electronic Science and Technology of China, Chengdu 611731, China (E-mail: xiuyue@std.uestc.edu.cn, Zhangzp, wn@uestc.edu.cn). Jun Zhao is with School of Computer Science and Engineering, Nanyang Technological University, Singapore (E-mail: junzhao@ntu.edu.sg).
Wei Sun is with School of Computer Science and Engineering, Northeastern University, Shenyang 110819, China (E-mail: weisun@stumail.neu.edu.cn).
Marco Di Renzo is with Universit\'e Paris-Saclay, CNRS, CentraleSup\'elec, Laboratoire des Signaux et Syst\`emes, 91192 Gif-sur-Yvette, France. (E-mail: marco.direnzo@centralesupelec.fr). 
Guan Gui is with College of Telecommunications and Information Engineering, Nanjing University of Posts and Telecommunications, Nanjing 210003, China (E-mail: guiguan@njupt.edu.cn).
}

\thanks{This work was supported in part by the Guangdong province Key Project of science and Technology (2018B010115001), the National Natural Science Foundation of China (NSFC) under Grant 91938202 and 61871070, and the Defense Industrial Technology Development Program (JCKY2016204A603). The corresponding author is Ning Wei.}}

\maketitle

\begin{abstract}
In this paper,  an reconfigurable intelligent surfaces (RIS)-aided millimeter wave (mmWave) non-orthogonal multiple access (NOMA) system is considered. In particular, we consider an RIS-aided mmWave-NOMA downlink system with a hybrid beamforming structure.  To maximize the achievable sum-rate under a minimum rate constraint for the users and a minimum transmit power constraint, a joint RIS phase shifts, hybrid beamforming, and power allocation problem is formulated. To solve this non-convex optimization problem, we develop an alternating optimization algorithm.  Specifically, first, the non-convex problem is transformed into three subproblems, i.e., power allocation, joint phase shifts and analog beamforming optimization, and digital beamforming design.  Then, we solve the power allocation problem under fixed phase shifts of the RIS and hybrid beamforming. Finally, given the power allocation matrix, an alternating manifold optimization (AMO)-based method and a successive convex approximation (SCA)-based method are utilized to design the phase shifts, analog beamforming, and transmit beamforming, respectively. Numerical results reveal that the proposed alternating optimization algorithm outperforms state-of-the-art schemes in terms of sum-rate. Moreover, compared to a conventional mmWave-NOMA system without RIS,  the proposed RIS-aided mmWave-NOMA system is capable of improving the achievable sum-rate of the system.
\end{abstract}


%
\IEEEpeerreviewmaketitle

\section{Introduction}
\IEEEPARstart{M}{illimeter} wave (mmWave) communications have been proposed as one of the candidate key {}{technologies} for the fifth-generation (5G) wireless {}{systems} and beyond ~\cite{roh2014millimeter, wang2015multi, di2020smart}. {}{In this context}, massive connectivity is a typical {}{requirement for several applications}. {}{In conventional systems, however, the data streams are transmitted, in each resource block, by employing an orthogonal multiple access (OMA) scheme\cite{dai2018hybrid}.} 
For mmWave communications with the OMA scheme,  the number of the users for each data stream in the same time-frequency-code-space resource block is one. Therefore, the total number of served
users is limited, which is no greater than the number of RF chains in each resource block \cite{xiao2018joint}. In \cite{dai2018survey}, {}{L. Dai et al.} proposed a mmWave non-orthogonal multiple access (NOMA) communication system {}{in order to overcome this issue}. {}{Based on this proposal, the signals are transmitted, in each resource block, by using power domain NOMA}.  In addition, the users {}{experiencing} different channel conditions {}{are} served simultaneously by employing superposition coding at the transmitter and successive interference cancellation (SIC) at the receiver\cite{liu2018deep}. {}{This approach} can greatly improve the number of served users.

Although mmWave-NOMA has many advantages in {}{terms of} improving {}{the} communications performance, {}{some limitations may prevent the potential application of} mmWave-NOMA.  Compared with conventional low-frequency communications, a key challenge {}{of} mmWave-NOMA {}{communications} is that the transmit signal usually suffers a severe path loss~\cite{alkhateeb2015limited}. Furthermore, {}{the use of highly directive antennas} makes mmWave-NOMA {}{communications} vulnerable to blockages. Finally, strong user interference {}{may limit} the application of mmWave-NOMA. Recently, reconfigurable intelligent surfaces {}{(RISs)} {}{have} been  proposed as a promising technology to {}{alleviate and possibly counteract} these problems~\cite{wu2019towards,guan2020joint, wu2019intelligent, guo2019weighted}. {}{An} RIS is a planar array comprising of a large number of reconfigurable  passive elements, which can reflect the incident signal {}{by appropriately tuning its amplitude and phase}. Therefore, {}{RISs have the capability of} {}{enhancing} the received signal power {}{and} suppress the co-channel interference {}{of} the users, {}{as well as} overcoming the path loss and {}{signals'} blockage {}{of} mmWave communications {}{thus} {}{making the transfer of information more reliable}. {}{In} conventional mmWave-NOMA, {}{in addition}, the decoding order is determined by the users’ channel power gains. {}{By using RISs}, the users’ decoding order can be designed {}{in a more flexible manner} by reconfiguring the RIS {}{phase shifts}, which introduces {}{additional} degrees-of-freedom (DoF) for {}{improving the performance of mmWave-NOMA systems.}

\subsection{Related Work}
{}{Thanks to the many potential benefits}, {}{RISs} have been investigated {}{for application to}
various wireless communication systems. {}{In} \cite{cao2019intelligent}, the joint power control and phase shift optimization problem {}{was} studied for {}{application to} mobile edge computing in RIS-aided mmWave systems. {}{Also}, a distributed optimization algorithm was proposed to solve the joint optimization problem. In \cite{jamali2019intelligent}, an architecture {}{for} RIS-aided mmWave massive multiple-input multiple-output (MIMO) {}{systems} was designed, and two efficient precoders were proposed by exploiting the sparsity of mmWave channels. {}{The design of hybrid analog-digital} precoding and phase shift optimization for RIS-aided mmWave {}{systems was} investigated in \cite{pradhan2020hybrid}, and an iterative algorithm was proposed to minimize the mean-squared-error (MSE). In \cite{wang2019intelligent}, the joint transmit beamforming and phase shift optimization problem was studied for multi-RIS-aided mmWave systems. For {}{application to} multiple-input single-output (MISO) RIS-aided NOMA systems, the semidefinite relaxation method and {}{the} manifold optimization method were used to solve the joint transmit beamforming and phase shift optimization problem \cite{zhu2019power}. A theoretical performance comparison between RIS-NOMA and RIS-OMA was
provided in \cite{zheng2020intelligent}, and a low-complexity algorithm was proposed {}{for achieving} near-optimal performance. The resource allocation
problem for a multi-channel RIS-aided NOMA system was studied
in \cite{zuo2020resource}, and an algorithm was proposed to jointly optimize the
subcarrier assignment, power allocation, and phase shifts. An
RIS-aided uplink NOMA system was considered in \cite{zeng2020sum}, and a near-optimal solution was proposed {}{for jointly optimizing} the {}{phase shifts} and {}{the transmit power}. Furthermore, many other research problems in the context of RIS-aided wireless communications have been recently addressed in the literature, which include information  rate  maximization  in \cite{zhou2020robust,qian2020beamforming,karasik2019beyond,perovic2019channel},  channel estimation in \cite{liu2020deep}, and robust optimization in \cite{han2019intelligent,zhou2020spectral}

\subsection{Motivations and Contributions}
Although these papers studied sum-rate enhancement for {}{NOMA-aided} wireless communication systems or {}{RIS-aided} mmWave communication systems, {}{none of them addressed the analysis and optimization of} RIS-aided mmWave-NOMA {}{systems} with a hybrid beamforming structure. Moreover, {}{the optimal power allocation and sum-rate maximization in} RIS-aided mmWave-NOMA system are challenging tasks to be tackled \cite{liu2018deep}, \cite{ning2019energy}. Motivated by {}{these considerations}, we investigate {}{an} RIS-aided mmWave-NOMA system with {}{a} hybrid beamforming structures{}{, and provide the following technical contributions}:
\begin{itemize}
\item
To maximize the sum-rate under {}{a} minimum rate {}{constraint for each user and a minimum transmit power constraint}, {}{we formulate} a joint optimization problem {}{for the transmit power}, the {}{phase shifts} of the RIS, and {}{the} hybrid beamforming. An alternating optimization algorithm is proposed to solve this problem.

\item
{}{By assuming that the phase shifts and the hybrid beamforming are fixed, we propose an algorithm for solving the power allocation problem.} Due to the nonconvexity of the {}{considered} problem, we divide it into two subproblems {}{and tackle both of them by applying alternating optimization methods}.

\item
We {}{optimize} the {}{phase shifts of the} RIS and {}{the} hybrid beamforming to suppress the interference while maximizing the sum-rate. In the proposed algorithm, the {}{phase shifts} and {}{the analog beamforming} are designed by using the alternating manifold optimization (AMO) algorithm and {}{by assuming that the transmit power and the digital beamforming weights are fixed}. {}{We} utilize the successive convex approximation (SCA)-based algorithm to solve the digital beamforming optimization problem. {}{ The convergence of the algorithms} is proved. 
\item
After optimizing the power allocation, the {}{phase shifts} of {}{the RIS}, and {}{the} hybrid beamforming, we evaluate the performance of the proposed 
algorithm for {}{application to} RIS-aided mmWave-NOMA {}{systems}. The numerical results reveal that the proposed RIS-aided mmWave-NOMA system {}{yields a better sum-rate than a}  traditional mmWave-NOMA system {}{that does not use RISs}. 
\end{itemize}

\textbf{Organization:} The rest of the paper is organized as follows. Section \ref{II} {}{introduces} the system model and problem {}{formulation}. Section \ref{III} reports the proposed power allocation algorithm.  In Section \ref{IV}, the AMO algorithm {}{for phase shifts} and hybrid beamforming optimization {}{is introduced}. Numerical results are {}{illustrated in Section \ref{V} in order} to evaluate the proposed alternating optimization algorithm. {}{Finally, 
S}ection \ref{VI} concludes the paper.

\textbf{Notation:} {}{The} imaginary unit is denoted by $j=\sqrt{-1}$. Matrices and vectors are denoted by boldface capital and lower-case letters, respectively. $\mathrm{diag}\{x_{1},\ldots,x_{N}\}$ denotes a diagonal matrix whose diagonal components are $x_{1},\ldots,x_{N}$. The real and imaginary parts of a complex number $x$ are denoted by $\mathrm{Re}(x)$ and $\mathrm{Im}(x)$, respectively. $\boldsymbol{x}^{*}$, $\boldsymbol{x}^{T}$, and $\boldsymbol{x}^{H}$ denote the conjugate, transpose, and conjugate transpose of vector $\boldsymbol{x}${}{, respectively}.  $x_{n}$ and $X_{k,n}$ {}{denote} the $n$th and $(k,n)$th elements of vector $\boldsymbol{x}$ and matrix $\boldsymbol{X}${}{, respectively}. $\|\boldsymbol{x}\|$ denotes the 2-norm of vector $\boldsymbol{x}$. $\mathcal{CN}(x,\sigma^{2})$ denotes the Gaussian distribution, {}{where} $x$ and $\sigma$ are the mean and variance, respectively.

\section{System Model and Problem Formulation}\label{II}
\begin{figure*}[h]
\centering
\includegraphics[scale=0.6]{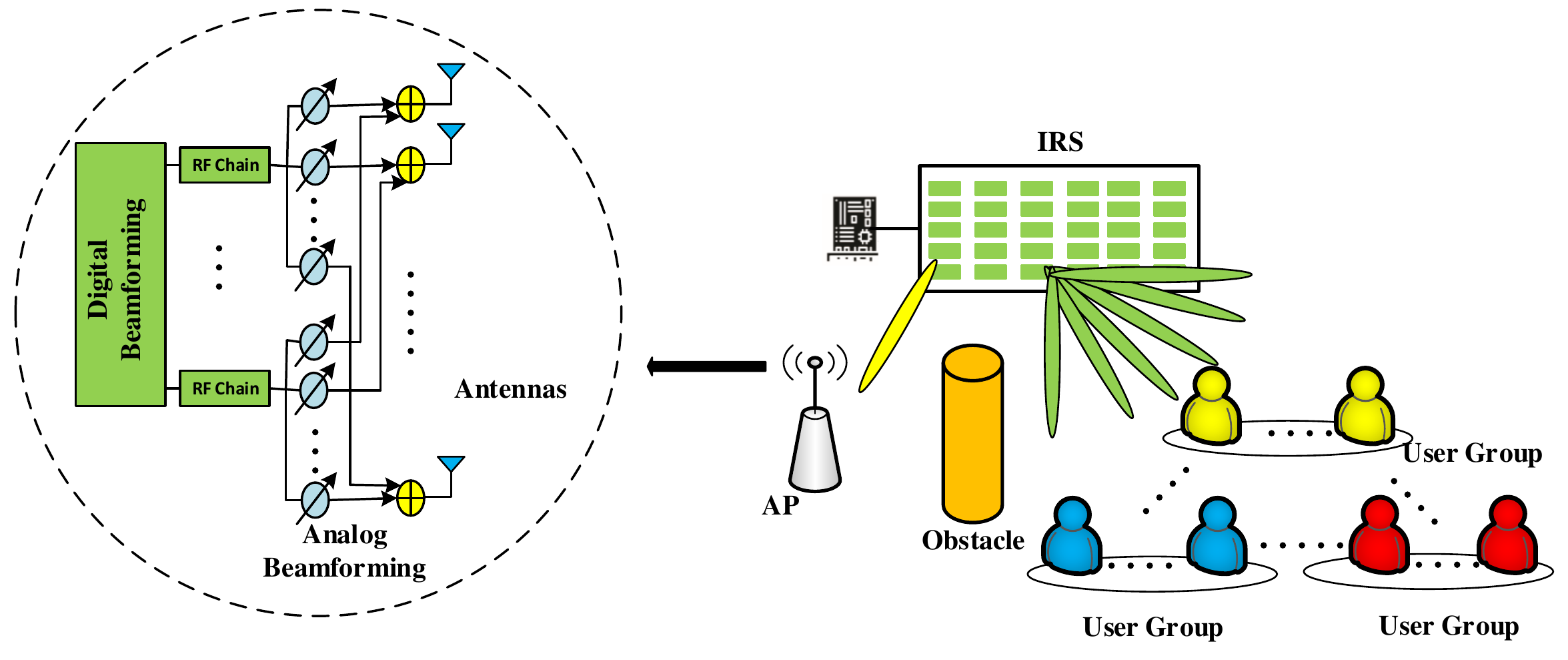}
  \caption{Illustration of the downlink multi-group multi-user RIS-aided mmWave-NOMA communication system.}
 \label{fig:3_1}
\end{figure*}
\newcounter{mytempeqncnt}

As shown in Fig.\ref{fig:3_1}, an RIS-aided mmWave-NOMA system is considered. The access point (AP) is equipped with a hybrid beamforming structure, where the number of transmit antennas and radio frequency (RF) chains are $N_{t}$ and $N_{RF}$, respectively.  $K$ users {}{equipped with} a single antenna {}{are distributed} in $N$ groups, {}{with} $K>N_{RF}$. Let $\boldsymbol{s}\in\mathbb{C}^{K\times 1}$ be the {}{transmitted} signals, {}{where} $\mathbb{E}(\boldsymbol{s}\boldsymbol{s}^{H})=\mathbf{I}$. The data streams are precoded by the digital beamforming matrix $\boldsymbol{W}\in\mathbb{C}^{N_{RF}\times N_{s}}$, where $N_{s}$ is the length of {}{each} data stream. Then, {}{an} analog beamforming {}{matrix} $\boldsymbol{F}\in\mathbb{C}^{N_{t}\times N_{RF}}$ {}{is applied. Analog beamforming is realized by using $N_t$ phase shifters.} {}{The} number of data streams is assumed to be equal to the number of RF chains.  In particular, each independent data stream corresponds to a group, i.e., $N=N_{s}=N_{RF}$. {}{The signals transmitted by the AP reach the RIS through the wireless channel}. {}{The RIS applies a phase shift matrix} $\boldsymbol{\Theta}=\mathrm{diag}(\boldsymbol{\theta})\in\mathbb{C}^{N_{r}\times N_{r}}$ to {}{the incident signals}, where $\boldsymbol{\theta}=[e^{j\theta_{1}},\cdots,e^{j\theta_{N_{r}}}]^{T}\in\mathbb{C}^{N_{r}\times 1}$, $0\leq \theta_{i}\leq 2\pi$, $1\leq i\leq N_{r}$ and $N_{r}$ is the number of reflecting elements {}{of the RIS}. The users perform successive interference cancellation (SIC) {}{on the signals reflected by the RIS. In particular, SIC is applied to users in the same group}. 
{}{We denote by} $\mathcal{G}_{n}$ the $n$th group, {}{which fulfills the properties} $\mathcal{G}_{i}\cap\mathcal{G}_{j}=\emptyset$, $\forall~i\neq j$, {}{where} $\emptyset$ {}{denotes} the empty set, $\sum_{n=1}^{N}|\mathcal{G}_{n}|=K$ {}{, and} $|\mathcal{G}_{n}|$ is the number of users {}{in} $\mathcal{G}_{n}$.  {}{In our system model, we assume} that  the  direct  link  between  the  AP  and  the user is  blocked, which  {}{is a typical application scenario when RISs are needed or used}\cite{sun2014mimo}. 
The AP-to-RIS channel is denoted {}{by} $\boldsymbol{G}\in\mathbb{C}^{N_{r}\times N_{t}}$, and the channel between the RIS and the user $k$ in the $n$th group {}{is denoted by} $\boldsymbol{h}_{n,k}\in\mathbb{C}^{N_{r}\times 1}$. Then, the received signal for the $k$th user in the $n$th group {}{can be written as follows}
\begin{eqnarray}
y_{n,k}=\boldsymbol{h}_{n,k}^{H}\boldsymbol{\Theta}\boldsymbol{G}\boldsymbol{F}\boldsymbol{W}\boldsymbol{P}\boldsymbol{s}+u_{n,k},\label{3-1}
\end{eqnarray}
where $u_{n,k}\sim\mathcal{CN}(0,\sigma^{2})$ is the noise at the user, and $\boldsymbol{P}=\mathrm{diag}\{\boldsymbol{p}_{1},\boldsymbol{p}_{2},\ldots,\boldsymbol{p}_{N}\}\in\mathbb{C}^{N\times K}$ is the power allocation matrix, where $\boldsymbol{p}_{n}=[\sqrt{p_{n,1}},\ldots,\sqrt{p_{n,|\mathcal{G}_{n}|}}]\in\mathbb{C}^{1\times|\mathcal{G}_{n}|}$. The analog beamforming matrix $\boldsymbol{F}$ 
and phase shift vector $\boldsymbol{\theta}$ with constant modulus constraint {}{are defined as follows} \cite{wu2019towards}
\begin{eqnarray}
|\boldsymbol{F}_{i,j}|=\frac{1}{\sqrt{N_{t}}},~1\leq i\leq N_{t}, 1\leq j\leq N_{RF},
|\boldsymbol{\theta}_{i}|=1,~1\leq i\leq N_{r}.\label{3-3}
\end{eqnarray}
{}{The} hybrid beamforming matrix {}{is defined} as
\begin{eqnarray}
\boldsymbol{D}=\boldsymbol{F}\boldsymbol{W}=[\boldsymbol{d}_{1},\boldsymbol{d}_{2},\ldots,\boldsymbol{d}_{N}].
\end{eqnarray}
Since the transmission power is separated from {}{the} hybrid beamforming matrix, each column of $\boldsymbol{D}$ {}{fulfills the property}
\begin{eqnarray}
\|\boldsymbol{d}_{n}\|=1,~1\leq n\leq N.\label{3-4}
\end{eqnarray}
Measurement campaigns {}{showed} that the power of the mmWave
{}{line-of-sight (LoS)} path is {}{usually} much higher (about $13$~dB higher) than the sum of the {}{powers} of non-line-of-sight (NLoS) paths\cite{akdeniz2014millimeter}. Considering this fact, it is desirable
to make sure that the channel between the AP and {}{the} RIS is {}{in LoS}. In practice, {}{by assuming that the} location of the AP is known, {}{the location of the RIS} can be appropriately {}{chosen}~{}{so as to ensure that the AP-RIS channel is in LoS}. {}{Based on these considerations, we assume that the} channel from the AP to the RIS can be well approximated by a rank-one matrix, i.e.,
\begin{eqnarray}
\boldsymbol{G}=\alpha\boldsymbol{a}_{r}(\phi)\boldsymbol{a}_{t}^{T}(\vartheta),\label{3-5}
\end{eqnarray}
where $\alpha$ is a scaling factor accounting for the antenna and path gains, where $\boldsymbol{a}_{t}(\vartheta)\in\mathbb{C}^{N_{t}\times 1}$ and $\boldsymbol{a}_{r}(\phi)\in\mathbb{C}^{N_{r}\times 1}$ represent the normalized array response {}{vectors} associated with the AP and the RIS, respectively. The channel from the RIS to the $k$th user in the $n$th group is
generated according to the following geometric channel model \cite{el2014spatially}
\begin{eqnarray}
\boldsymbol{h}_{n,k}=\sum_{l=0}^{L_{n,k}-1}\beta_{n,k}^{l}\boldsymbol{b}_{t}(\theta_{l}),\label{3-6}
\end{eqnarray}
where $\beta_{n,k}^{l}$ is a scaling factor accounting for the antenna and path gains, $\boldsymbol{b}_{t}(\theta_{l})\in\mathbb{C}^{N_{r}\times 1}$ represents the normalized array response vector of the RIS, and $L_{n,k}$ is the total number of paths.

{}{In power domain NOMA, in general}, the optimal decoding order of the {}{users} is determined based on the {}{users'} channel {}{gains} \cite{saito2013non}. {}{In an RIS-aided} mmWave-NOMA {}{system with} hybrid beamforming, {}{on the other hand}, the decoding order is determined by the channel gains and {}{by} the beamforming gains. Therefore, {}{it is necessary to} {}{first} determine the decoding order. Without loss of generality, we assume the decoding order in the $n$th group is
$|\boldsymbol{h}_{n,1}^{H}\boldsymbol{\Theta}\boldsymbol{G}\boldsymbol{F}\boldsymbol{w}_{n}|^{2}\geq|\boldsymbol{h}_{n,2}^{H}\boldsymbol{\Theta}\boldsymbol{G}\boldsymbol{F}\boldsymbol{w}_{n}|^{2}\geq\cdots\geq|\boldsymbol{h}_{n,|\mathcal{G}_{n}|}^{H}\boldsymbol{\Theta}\boldsymbol{G}\boldsymbol{F}\boldsymbol{w}_{n}|^{2}$,
which implies that the optimal decoding order is determined by the effective channel gains, {}{ranked in increasing order of magnitude}\cite{dai2018hybrid,xiao2018joint,ding2017survey}. Thus, the user $k$ in the group $n$ can decode $s_{n,k}$ $(n+1 \leq j\leq |\mathcal{G}_{n}|)$, {}{which is then} removed from the received signal. The other signals are treated as
interference. Therefore, the {}{signal to interference plus noise ratio (SINR)} of the user $k$ in group $n$ is given by
\begin{eqnarray}
\mathrm{SINR}_{n,k}=\frac{|\boldsymbol{h}_{n,k}^{H}\boldsymbol{\Theta}\boldsymbol{G}\boldsymbol{F}\boldsymbol{w}_{n}|^{2}p_{n,k}}{|\boldsymbol{h}_{n,k}^{H}\boldsymbol{\Theta}\boldsymbol{G}\boldsymbol{F}\boldsymbol{w}_{n}|^{2}\sum_{j=1}^{k-1}p_{n,j}+\sum_{i\neq n}\sum_{j=1}^{|\mathcal{G}_{i}|}|\boldsymbol{h}_{n,k}^{H}\boldsymbol{\Theta}\boldsymbol{G}\boldsymbol{F}\boldsymbol{w}_{i}|^{2}p_{i,j}+\sigma^{2}}.\label{3-7}
\end{eqnarray}
In addition, {}{we} assume {}{that} the interference from other groups can be well suppressed if the following holds
\begin{eqnarray}
|\boldsymbol{h}_{n,k}^{H}\boldsymbol{\Theta}\boldsymbol{G}\boldsymbol{F}\boldsymbol{w}_{i}|^{2}\leq\tau,~i\neq n,\label{3-8}
\end{eqnarray}
where $\tau$ is small enough. Therefore, the {}{main} interference {}{originates} from other users in the same group. {}{Under these assumptions} the $\mathrm{SINR}_{n,k}$ can be approximated as
\begin{eqnarray}
\mathrm{SINR}_{n,k}\approx\frac{|\boldsymbol{h}_{n,k}^{H}\boldsymbol{\Theta}\boldsymbol{G}\boldsymbol{F}\boldsymbol{w}_{n}|^{2}p_{n,k}}{|\boldsymbol{h}_{n,k}^{H}\boldsymbol{\Theta}\boldsymbol{G}\boldsymbol{F}\boldsymbol{w}_{n}|^{2}\sum_{j=1}^{k-1}p_{n,j}+\sigma^{2}}.\label{3-9}
\end{eqnarray}
According to (\ref{3-9}), the achievable rate of the user $k$ in group $n$ {}{can be written as}
\begin{eqnarray}
R_{n,k}=\log_{2}(1+\mathrm{SINR}_{n,k}).\label{3-10}
\end{eqnarray}
Finally, the sum-rate of the RIS-aided mmWave-NOMA system {}{can be written as}
\begin{eqnarray}
R=\sum_{n=1}^{N}\sum_{k=1}^{|\mathcal{G}_{n}|}R_{n,k}.\label{3-11}
\end{eqnarray}

We assume that the AP-to-RIS channel and RIS-to-user channel are known by the AP, and {}{that} all optimization operations are {}{executed} at the AP. Based  on  (\ref{3-9})-(\ref{3-11}),  the  sum-rate  optimization  problem  for  RIS-aided  mmWave-NOMA with hybrid beamforming {}{can be} formulated as
\begin{subequations}
\begin{align}
\max_{\boldsymbol{W},\boldsymbol{\Theta},\boldsymbol{F}, \{p_{n,k}\}}&R\label{3-12a}\\
\mbox{s.t.}~
&R_{n,k}\geq \gamma_{n,k}\label{3-12b}\\
&p_{n,k}\geq 0,&\label{3-12c}\\
&\sum_{n=1}^{N}\sum_{k=1}^{|\mathcal{G}_{n}|}p_{n,k}\leq P,&\label{3-12d}\\
&|\boldsymbol{F}_{i,j}|=\frac{1}{\sqrt{N}}, 1\leq i\leq N_{t}, 1\leq j\leq N_{RF},&\label{3-12e}\\
&|\boldsymbol{\theta}_{i}|=1, 1\leq i\leq N_{r},&\label{3-12f}\\
&\boldsymbol{D}=\boldsymbol{F}\boldsymbol{W},&\label{3-12g}\\
&\|\boldsymbol{d}_{n}\|=1, 1\leq n\leq N,&\label{3-12h}\\
&|\boldsymbol{h}_{n,k}^{H}\boldsymbol{\Theta}\boldsymbol{G}\boldsymbol{F}\boldsymbol{w}_{i}|^{2}\leq\tau,~i\neq n.&\label{3-12i}
\end{align}\label{3-12}%
\end{subequations}

{}{The size} of all variables in the problem (\ref{3-12}) {}{is} $N_{RF}N_{s}+N_{RF}N_{t}+K+N_{r}$, {}{which is usually large}. {}{Two major challenges render the solution of the optimization problem in (\ref{3-12}) difficult to tackle.} The first difficulty is that the optimized variables are {}{coupled}, which makes the problem non-convex. The second difficulty is the decoding order. {}{Usually}, the optimal decoding order {}{corresponds to the} increasing order of the users’ effective channel gains. However, the {}{ordering} of {}{the} effective channel gains varies with {}{the} beamforming matrix and {}{the} phase shift matrix. These challenges make the problem in (\ref{3-12}) difficult to solve. To {}{tackle both issues}, an alternating optimization algorithm is proposed, which includes three parts, i.e., power optimization, analog beamforming and RIS {}{phase shifts} optimization, and transmit beamforming optimization. {}{These three sub-parts are analyzed in the following sections}.
 

\section{Power Allocation Optimization}\label{III}
{}{For a given} hybrid beamforming matrix and a {}{given} {}{phase shifts} matrix of {}{the} RIS, {}{(\ref{3-12})} can be simplified as
\begin{subequations}
\begin{align}
\max_{\{p_{n,k}\}}&~R\label{3-13a}\\
\mbox{s.t.}~
&\text{(\ref{3-12b})},\text{(\ref{3-12c})},\text{(\ref{3-12d})}.&\label{3-13b}
\end{align}\label{3-13}%
\end{subequations}
{}{When} the hybrid beamforming matrix and {}{the phase shifts} matrix of the RIS are fixed, the decoding order is fixed. {}{This simplifies the optimization problem to solve.} Without loss of generality, we {}{assume} $|\boldsymbol{h}_{n,1}^{H}\boldsymbol{\Theta}\boldsymbol{G}\boldsymbol{F}\boldsymbol{w}_{n}|^{2}\geq|\boldsymbol{h}_{n,2}^{H}\boldsymbol{\Theta}\boldsymbol{G}\boldsymbol{F}\boldsymbol{w}_{n}|^{2}\geq\cdots\geq|\boldsymbol{h}_{n,|\mathcal{G}_{n}|}^{H}\boldsymbol{\Theta}\boldsymbol{G}\boldsymbol{F}\boldsymbol{w}_{n}|^{2}$. Although the decoding order is fixed, the
objective function and the constraint (\ref{3-12b}) in (\ref{3-12}) are still non-convex. To address this
{}{issue}, we introduce the auxiliary variables $\{P_{n}\}$ and $P_{n}=\sum_{k=1}^{|\mathcal{G}_{n}|}p_{n,k}$, $\forall i\leq n\leq N$, which denotes the {}{power allocated to} the $n$th group. Therefore, (\ref{3-13}) is reformulated as
\begin{subequations}
\begin{align}
\max_{\{P_{n}\}}\max_{\{p_{n,k}\}}&~R\label{3-15a}\\
\mbox{s.t.}~
&\text{(\ref{3-12b})},\text{(\ref{3-12c})},&\label{3-15b}\\
&\sum_{k=1}^{|\mathcal{G}_{n}|}p_{n,k}=P_{n},&\label{3-15c}\\
&\sum_{n=1}^{N}P_{n}=P.&\label{3-15d}
\end{align}\label{3-15}%
\end{subequations}
Since we assume that {}{the} interference from other groups {}{can be} suppressed, according to (\ref{3-9}), {}{a user in} group $n$ {}{is} mainly interfered by other users in group $n$.
{}{Based on the considered decoding order and taking into account the results in the results in} \cite{zhu2019joint}, {}{the} maximization {}{of the} sum-rate {}{corresponds to setting}, $R_{n,k}=\gamma_{k}$, $\forall~1\leq n\leq N$, $2\leq k\leq |\mathcal{G}_{n}|$, i.e., {}{the user with the higher decoding order are allocated as much power as possible if the other users meet the minimum rate requirements.} Hence, the {}{solution of the} power allocation problem in group $n$ {}{corresponds to the following equations}
\begin{eqnarray}
R_{n,2}=\gamma_{n,2},~\ldots,
~R_{n,|\mathcal{G}_{n}|}&=&\gamma_{n,|\mathcal{G}_{n}|},~\sum_{k=1}^{|\mathcal{G}_{n}|}p_{n,k}=P_{n}.\label{3-16}
\end{eqnarray}
{}{From (11) and (12)}, the solution of equation (\ref{3-16}) {}{is}
\begin{align}
&p_{n,|\mathcal{G}_{n}|}=\frac{2^{\gamma_{n,|\mathcal{G}_{n}|}}}{1+2^{\gamma_{n},|\mathcal{G}_{n}|}}\left(P_{n}+\frac{\sigma^{2}}{|\boldsymbol{h}_{n,|\mathcal{G}_{n}|}\boldsymbol{\Theta}\boldsymbol{G}\boldsymbol{F}\boldsymbol{w}_{n}|^{2}}\right),\nonumber\\
&~~~~~~~~~~~\vdots\nonumber\\
&p_{n,2}=\frac{2^{\gamma_{n,2}}}{1+2^{\gamma_{n},2}}\left(P_{n}-\sum_{k=3}^{|\mathcal{G}_{n}|}p_{n,k}+\frac{\sigma^{2}}{|\boldsymbol{h}_{n,2}\boldsymbol{\Theta}\boldsymbol{G}\boldsymbol{F}\boldsymbol{w}_{n}|^{2}}\right),\nonumber\\
&p_{n,1}=P_{n}-\sum_{k=2}^{|\mathcal{G}_{n}|}p_{n,k}.\label{3-17}
\end{align}
{}{Since} $R_{n,2}=\gamma_{n,2},\cdots,R_{n,|\mathcal{G}_{n}|}=\gamma_{n,|\mathcal{G}_{n}|}$, the objective function in (\ref{3-15a}) can be rewritten as
\begin{eqnarray}
R=\sum_{n=1}^{N}R_{n,1}+\sum_{n=1}^{N}\sum_{k=2}^{|\mathcal{G}_{n}|}\gamma_{n,k}.\label{3-18}
\end{eqnarray}
Since $\sum_{n=1}^{N}\sum_{k=2}^{|\mathcal{G}_{n}|}\gamma_{n,k}$ is a constant, (\ref{3-15}) is further simplified as 
\begin{subequations}
\begin{align}
\max_{\{P_{n}\}}&\sum_{n=1}^{N}R_{n,1},\label{3-19a}\\
\mbox{s.t.}~
&\text{(\ref{3-12b})},\text{(\ref{3-15d})}.&\label{3-19b}
\end{align}\label{3-19}%
\end{subequations}
Because the objective function (\ref{3-19a}) is non-convex, solving problem (\ref{3-19}) is still difficult. We propose an iterative algorithm {}{to tackle it}. 
{}{To this end, we note that the optimization problem in} (\ref{3-19}) without (\ref{3-12d}) is convex. Hence, we have the following \textbf{Theorem~\ref{The31}}.
\begin{theorem}\label{The31}
When the constraint (\ref{3-12b}) is removed, the globally optimal solution of (\ref{3-19}) is
\begin{eqnarray}
\bar{P}_{n}=\frac{P}{N}-\frac{\alpha_{n}+1}{\beta_{n}}+\sum_{i=1}^{N}\frac{\alpha_{i}+1}{N\beta_{i}},\label{3-26}
\end{eqnarray}
where 
\begin{align}
\beta_{n}=\frac{|\boldsymbol{h}_{n,1}^{H}\boldsymbol{\Theta}\boldsymbol{G}\boldsymbol{F}\boldsymbol{w}_{n}|^{2}}{\sigma^{2}}\left(1-\sum_{k=2}^{|\mathcal{G}_{n}|}\left[(2^{\gamma_{n,k}}-1)\prod_{j=2}^{k}\frac{1}{2^{\gamma_{n,j}}}\right]\right)
\end{align}
and
\begin{align}
\alpha_{n}=-\frac{|\boldsymbol{h}_{n,1}^{H}\boldsymbol{\Theta}\boldsymbol{G}\boldsymbol{F}\boldsymbol{w}_{n}|^{2}}{\sigma^{2}}\sum_{k=2}^{|\mathcal{G}_{n}|}\left[(2^{\gamma_{n,k}}-1)\frac{\sigma^{2}}{|\boldsymbol{h}_{n,k}^{H}\boldsymbol{\Theta}\boldsymbol{G}\boldsymbol{F}\boldsymbol{w}_{k}|^{2}}\prod_{j=2}^{k}\frac{1}{2^{\gamma_{n,j}}}\right].
\end{align}
The proof is given in \textbf{Appendix~A}.
\end{theorem}
Based on \textbf{Throrem}~\ref{The31}, if $\bar{P}_{n}$ in (\ref{3-26}) {}{satisfies} the constraint in (\ref{3-12b}), i.e., $\frac{|\boldsymbol{h}_{n,k}^{H}\boldsymbol{\Theta}\boldsymbol{G}\boldsymbol{F}\boldsymbol{w}_{n}|^{2}p_{n,k}}{|\boldsymbol{h}_{n,k}^{H}\boldsymbol{\Theta}\boldsymbol{G}\boldsymbol{F}\boldsymbol{w}_{n}|^{2}(\bar{P}_{n}-p_{n,k})+\sigma^{2}}\geq 2^{\gamma_{n,k}}-1$, $\bar{P_{n}}$ is the optimal solution of (\ref{3-19}). However, if $\bar{P_{n}}$ do not satisfy the constraint in (\ref{3-12b}), i.e., $\frac{|\boldsymbol{h}_{n,k}^{H}\boldsymbol{\Theta}\boldsymbol{G}\boldsymbol{F}\boldsymbol{w}_{n}|^{2}p_{n,k}}{|\boldsymbol{h}_{n,k}^{H}\boldsymbol{\Theta}\boldsymbol{G}\boldsymbol{F}\boldsymbol{w}_{n}|^{2}(\bar{P}_{n}-p_{n,k})+\sigma^{2}}\leq 2^{\gamma_{n,k}}-1$, $\bar{P}_{n}$ is not the optimal solution of (\ref{3-19}). {}{In this latter case, the result in \textbf{Theorem \ref{The32}} can be used.} 
\begin{theorem}\label{The32}
{When the constraint (\ref{3-12b}) is considered, the globally optimal solution should always satisfy}
\begin{eqnarray}
\hat{P}_{n}=\frac{2^{\gamma_{n,1}}-\alpha_{n}}{\beta_{n}},~\forall~n\in\left\{n|1\leq n\leq N,\frac{|\boldsymbol{h}_{n,k}^{H}\boldsymbol{\Theta}\boldsymbol{G}\boldsymbol{F}\boldsymbol{w}_{n}|^{2}p_{n,k}}{|\boldsymbol{h}_{n,k}^{H}\boldsymbol{\Theta}\boldsymbol{G}\boldsymbol{F}\boldsymbol{w}_{n}|^{2}(\bar{P}_{n}-p_{n,k})+\sigma^{2}}< 2^{\gamma_{n,k}}-1\right\}.\label{3-27}
\end{eqnarray}

The proof is given in \textbf{Appendix~B}.
\end{theorem}
When $n\in\left\{n|1\leq n\leq N,\frac{|\boldsymbol{h}_{n,k}^{H}\boldsymbol{\Theta}\boldsymbol{G}\boldsymbol{F}\boldsymbol{w}_{n}|^{2}p_{n,k}}{|\boldsymbol{h}_{n,k}^{H}\boldsymbol{\Theta}\boldsymbol{G}\boldsymbol{F}\boldsymbol{w}_{n}|^{2}(\bar{P}_{n}-p_{n,k})+\sigma^{2}}\leq 2^{\gamma_{n,k}}-1\right\}$, the optimal
power allocation can be obtained by solving the following problem
\begin{subequations}
\begin{align}
\max_{\{P_{n}\}}&\sum_{n\notin\mathcal{N}}R_{n,1}\label{3-28a}\\
\mbox{s.t.}~
&R_{n,1}\geq r_{n,1},&\label{3-28b}\\
&\sum_{n\notin\mathcal{N}}P_{n}\leq P-\sum_{j\in\mathcal{N}}\hat{P}_{j},&\label{3-28c}
\end{align}\label{3-28}%
\end{subequations}
where $\mathcal{N}=\left\{n|1\leq n\leq N,\frac{|\boldsymbol{h}_{n,k}^{H}\boldsymbol{\Theta}\boldsymbol{G}\boldsymbol{F}\boldsymbol{w}_{n}|^{2}p_{n,k}}{|\boldsymbol{h}_{n,k}^{H}\boldsymbol{\Theta}\boldsymbol{G}\boldsymbol{F}\boldsymbol{w}_{n}|^{2}(\bar{P}_{n}-p_{n,k})+\sigma^{2}}\leq 2^{\gamma_{n,k}}-1\right\}$. Based on (\ref{3-28}), the proposed algorithms consists of the following steps. First, we consider the problem in (\ref{3-28}) by ignoring the constraint (\ref{3-28b}). In this case, \textbf{Theorem \ref{The31}} can be used. Then, we use \textbf{Theorem \ref{The32}} in order to obtain the solutions that do not satisfy the constraint (\ref{3-28b}), and update problem (\ref{3-28}). This procedure is iterated until convergence, as reported in \textbf{Algorithm}~\ref{algo3-1}.
\begin{algorithm}%
\caption{Proposed Group Power Allocation Algorithm} \label{algo3-1}
\hspace*{0.02in}{\bf Initialization:}
$t=0$, $P_{n}^{(t)}=P_{n}^{(0)}=\frac{P}{N_{RF}}$.\\
\hspace*{0.02in}{\bf Repeat:}\\
$\mathcal{V}=\{1,2,\cdots,N\}$, let $\mathcal{N}=\mathcal{V}$.\\
\hspace*{0.02in}{\bf If:}~$\mathcal{N}\neq\emptyset$\\
\hspace*{0.02in}{\bf Repeat:}\\
According to \textbf{Theorem}~\ref{The31}, calculate $\beta_{n}$, $\alpha_{n}$, and $\bar{P}_{n}^{(t)}$.\\
Update $\mathcal{N}=\left\{n|1\leq n\leq N,\frac{|\boldsymbol{h}_{n,k}^{H}\boldsymbol{\Theta}\boldsymbol{G}\boldsymbol{F}\boldsymbol{w}_{n}|^{2}p_{n,k}}{|\boldsymbol{h}_{n,k}^{H}\boldsymbol{\Theta}\boldsymbol{G}\boldsymbol{F}\boldsymbol{w}_{n}|^{2}(\bar{P}_{n}-p_{n,k})+\sigma^{2}}\leq 2^{\gamma_{n,k}}-1\right\}$.\\
According to \textbf{Theorem}~\ref{The32}, calculate $\hat{P}_{n}^{(t)}$.\\
Update the set according to $\mathcal{V}=\mathcal{V}/\mathcal{N}$.\\
\hspace*{0.02in}{\bf Until:}
$\mathcal{N}=\emptyset$.\\
\hspace*{0.02in}{\bf Update:} $\hat{P}_{n}^{(t)}=\bar{P}_{n}^{(t)}$.\\
\hspace*{0.02in}{\bf Until:}
$P_{n}^{*}=\hat{P}_{n}^{(T_{max})}$, where $T_{max}$ is maximize the number of iterations.\\
\hspace*{0.02in}{\bf Output:}
$\{P_{n}^{*}\}$.\\
\end{algorithm}

\section{Phase Shifts and Hybrid Beamforming Design}\label{IV}%
Given $\{p_{n,k}\}$ and $\{P_{n}\}$, the original problem {}{in} (\ref{3-12}) can be simplified as {}{follows}
\begin{subequations}
\begin{align}
\max_{\boldsymbol{\Theta},\boldsymbol{F},\boldsymbol{W}}&~R\label{3-14a}\\
\mbox{s.t.}~
&\text{(\ref{3-12b})},\text{(\ref{3-12e})},\text{(\ref{3-12f})},\text{(\ref{3-12g})},\text{(\ref{3-12h})},\text{(\ref{3-12i})}.\label{3-14b}
\end{align}\label{3-14}%
\end{subequations}
The non-convex modulus constraints for {}{the} analog beamforming and {}{the} {}{phase shifts of the RIS} make the {}{solving (\ref{3-14a}) difficult}. {}{To tackle (\ref{3-14a})}, we propose a suboptimal algorithm. First, we introduce the auxiliary variables $\{\boldsymbol{u}_{i}\}$, $\{v_{n,k,i}\}$, $\{z_{n,k,i}\}$, such that
\begin{eqnarray}
\boldsymbol{u}_{i}=\boldsymbol{G}\boldsymbol{F}\boldsymbol{w}_{i},
v_{n,k,i}=\boldsymbol{h}_{n,k}^{H}\boldsymbol{\Theta}\boldsymbol{u}_{i},
z_{n,k,i}=v_{n,k,i}v_{n,k,i}^{H}.\label{3-31}
\end{eqnarray}
Substituting (\ref{3-31}) into (\ref{3-12a}), the problem {}{in} (\ref{3-12}) can be rewritten as
\begin{subequations}
\begin{align}
\max_{\boldsymbol{W},\boldsymbol{\Theta},\boldsymbol{F},\{\boldsymbol{u}_{i}\},\{v_{n,k,i}\},\{z_{n,k,i}\}}&\sum_{n=1}^{N}\sum_{k=1}^{|\mathcal{G}_{n}|}\log_{2}\left(1+\frac{z_{n,k,n}p_{n,k}}{z_{n,k,n}\sum_{j=1}^{k-1}p_{n,j}+\sigma^{2}}\right)\label{3-32a}\\
\mbox{s.t.}~
&z_{n,k,n}p_{n,k}-(2^{\gamma_{n,k}}-1)(\boldsymbol{z}_{n,k,n}\sum_{j=1}^{k-1}p_{n,j}+\sigma^{2})\leq 0&\label{3-32b}\\
&\text{(\ref{3-12e})},\text{(\ref{3-12f})},\text{(\ref{3-12g})},\text{(\ref{3-12h})},&\label{3-32c}\\
&\boldsymbol{u}_{i}=\boldsymbol{G}\boldsymbol{F}\boldsymbol{w}_{i},\label{3-32d}&\\
&v_{n,k,i}=\boldsymbol{h}_{n,k}^{H}\boldsymbol{\Theta}\boldsymbol{u}_{i},\label{3-32e}&\\
&z_{n,k,i}=v_{n,k,i}v_{n,k,i}^{H},\label{3-32f}&\\
&z_{n,k,i}\leq\tau,i\neq n.\label{3-32g}&
\end{align}\label{3-32}%
\end{subequations}
{}{In order to} transform the objective function in (\ref{3-32a}) {}{into a difference of convex (DC)} programming problem, {}{we note that the objective function}~(\ref{3-32a}) {}{is equal to}
\begin{align}
&\sum_{n=1}^{N}\sum_{k=1}^{|\mathcal{G}_{n}|}\log_{2}(z_{n,k,n}\sum_{j=1}^{k}p_{n,j}+\sigma^{2})-\sum_{n=1}^{N}\sum_{k=1}^{|\mathcal{G}_{n}|}\log_{2}(z_{n,k,n}\sum_{j=1}^{k-1}p_{n,j}+\sigma^{2}).\label{3-33}
\end{align}
According to \cite{boyd2004convex}, the problem in (\ref{3-32}) is equivalent to
\begin{subequations}
\begin{align}
\min_{\boldsymbol{W},\boldsymbol{\Theta},\boldsymbol{F},\{\boldsymbol{u}_{i}\},\{v_{n,k,i}\},\{z_{n,k,i}\}}&\sum_{n=1}^{N}\sum_{k=1}^{|\mathcal{G}_{n}|}\log_{2}(z_{n,k,n}\sum_{j=1}^{k-1}p_{n,j}+\sigma^{2})-\sum_{n=1}^{N}&\sum_{k=1}^{|\mathcal{G}_{n}|}\log_{2}(z_{n,k,n}\sum_{j=1}^{k}p_{n,j}+\sigma^{2})\label{3-34a}\\
\mbox{s.t.}~
&\text{(\ref{3-12e})},\text{(\ref{3-12f})},\text{(\ref{3-12g})},\text{(\ref{3-12h})},\text{(\ref{3-32b})}, \nonumber \\
&\text{(\ref{3-32d})}, \text{(\ref{3-32e})},\text{(\ref{3-32f})},\text{(\ref{3-32g})}.&\label{3-34b}
\end{align}\label{3-34}%
\end{subequations}
Now, we focus on {}{the} constraint (\ref{3-32f}). The following theorem holds.

\begin{theorem}\label{The33}
Constraint (\ref{3-32f}) is equivalent to the following
\begin{eqnarray}
&\left[
 \begin{matrix}
   z_{n,k,i} & v_{n,k,i} \\
  v_{n,k,i}^{H} & 1\\
  \end{matrix}
  \right]\succeq\boldsymbol{0},&\label{3-37}
\end{eqnarray}
and
\begin{eqnarray}
z_{n,k,i}-v_{n,k,i}v_{n,k,i}^{H}\leq 0.\label{3-38}
\end{eqnarray}
The proof is given in \textbf{Appendix~C}.
\end{theorem}
Substituting (\ref{3-37}) and (\ref{3-38}) into (\ref{3-34}), {}{the} problem in (\ref{3-34}) is rewritten as
\begin{subequations}
\begin{align}
\min_{\boldsymbol{W},\boldsymbol{\Theta},\boldsymbol{F},\{\boldsymbol{u}_{i}\},\{v_{n,k,i}\},\{z_{n,k,i}\}}&\sum_{n=1}^{N}\sum_{k=1}^{|\mathcal{G}_{n}|}\log_{2}(z_{n,k,n}\sum_{j=1}^{k-1}p_{n,j}+\sigma^{2})
-\sum_{n=1}^{N}\sum_{k=1}^{|\mathcal{G}_{n}|}\log_{2}(z_{n,k,n}\sum_{j=1}^{k}p_{n,j}+\sigma^{2})\label{3-37a}\\
\mbox{s.t.}~
&\text{(\ref{3-12e})},\text{(\ref{3-12f})},\text{(\ref{3-12g})},\text{(\ref{3-12h})},\text{(\ref{3-32b})},&\nonumber\\
&\text{(\ref{3-32d})},\text{(\ref{3-32e})},\text{(\ref{3-32g})}&\label{3-37b}\\
&(z_{n,k,i}-v_{n,k,i}v_{n,k,i}^{H})\leq 0,&\label{3-37c}\\
&\left[
 \begin{matrix}
   z_{n,k,i} & v_{n,k,i} \\
  v_{n,k,i}^{H} & 1\\
  \end{matrix}
  \right]\succeq\boldsymbol{0}.&\label{3-37d}
\end{align}\label{3-37m}%
\end{subequations}
Employing  the  exact  penalty  method  \cite{boyd2004convex},  (\ref{3-37m})  can be rewritten as
\begin{subequations}
\begin{align}
\min_{\boldsymbol{W},\boldsymbol{\Theta},\boldsymbol{F},\{\boldsymbol{u}_{i}\},\{v_{n,k,i}\},\{z_{n,k,i}\}}&\sum_{n=1}^{N}\sum_{k=1}^{|\mathcal{G}_{n}|}\log_{2}(z_{n,k,n}\sum_{j=1}^{k-1}p_{n,j}+\sigma^{2})-\sum_{n=1}^{N}\sum_{k=1}^{|\mathcal{G}_{n}|}\log_{2}(z_{n,k,n}\sum_{j=1}^{k}p_{n,j}+\sigma^{2})\nonumber\\
+\lambda\Big(\sum_{i=1}^{N}&\|\boldsymbol{u}_{i}-\boldsymbol{G}\boldsymbol{F}\boldsymbol{w}_{i}\|^{2}+\|\boldsymbol{D}-\boldsymbol{F}\boldsymbol{W}\|^{2}
+\sum_{n=1}^{N}\sum_{k=1}^{|\mathcal{G}_{n}|}\sum_{i=1}^{N}\|v_{n,k,i}-\boldsymbol{h}_{n,k}^{H}\boldsymbol{\Theta}\boldsymbol{u}_{i}\|^{2}+\sum_{n=1}^{N}\nonumber\\
\sum_{k=1}^{|\mathcal{G}_{n}|}&\sum_{i=1}^{N}(z_{n,k,i}-v_{n,k,i}v_{n,k,i}^{H})\Big)\label{3-38a}\\
\mbox{s.t.}~
&\text{(\ref{3-12e})},\text{(\ref{3-12f})},\text{(\ref{3-12g})},\text{(\ref{3-12h})},\text{(\ref{3-32b})},&\nonumber\\
&\text{(\ref{3-32g})},\text{(\ref{3-37c})},\text{(\ref{3-37d})}.&\label{3-38b}
\end{align}\label{3-38m}%
\end{subequations}
{}{We observe that the minuend and the subtrahend in the} objective function and {}{the} constraints (\ref{3-32b}), (\ref{3-32g}), and (\ref{3-37d}) are convex, but the constraints (\ref{3-12e})-(\ref{3-12h}) and (\ref{3-37c}) are still non-convex. To deal with the non-convex {}{constraints} in (\ref{3-12e}), (\ref{3-12f}), and (\ref{3-12h}), {}{we propose an} AMO algorithm.

\subsection{Phase Shift and Analog Beamforming Design Based on {}{the} AMO Algorithm}
For {}{a} fixed transmit beamforming and power allocation matrix, problem (\ref{3-38m}) {}{simplifies to}
\begin{subequations}
\begin{align}
\min_{\boldsymbol{D},\boldsymbol{\Theta},\boldsymbol{F}}&\sum_{n=1}^{N}\sum_{k=1}^{|\mathcal{G}_{n}|}\log_{2}(z_{n,k,n}\sum_{j=1}^{k-1}p_{n,j}+\sigma^{2})-\sum_{n=1}^{N}\sum_{k=1}^{|\mathcal{G}_{n}|}\log_{2}(z_{n,k,n}\sum_{j=1}^{k}p_{n,j}+\sigma^{2})\nonumber\\
&+\lambda\Big(\sum_{i=1}^{N}\|\boldsymbol{u}_{i}-\boldsymbol{G}\boldsymbol{F}\boldsymbol{w}_{i}\|^{2}+\|\boldsymbol{D}-\boldsymbol{F}\boldsymbol{W}\|^{2}+\sum_{n=1}^{N}\sum_{k=1}^{|\mathcal{G}_{n}|}\sum_{i=1}^{N}\|v_{n,k,i}-\boldsymbol{h}_{n,k}^{H}\boldsymbol{\Theta}\boldsymbol{u}_{i}\|^{2}\nonumber\\
&+\sum_{n=1}^{N}\sum_{k=1}^{|\mathcal{G}_{n}|}\sum_{i=1}^{N}(z_{n,k,i}-v_{n,k,j}v_{n,k,i}^{H})\Big)\label{3-39a}\\
\mbox{s.t.}~
&\text{(\ref{3-12e})},\text{(\ref{3-12f})},\text{(\ref{3-12h})},&\label{3-39b}
\end{align}\label{3-39}%
\end{subequations}
According  to  the  notion  of manifold optimization, problem (\ref{3-39}) can be reformulated {}{in} three sub-problems:
\begin{align}
&\min_{\mathcal{M}_{1}}~\sum_{n=1}^{N}\sum_{k=1}^{|\mathcal{G}_{n}|}\sum_{i=1}^{N}\|v_{n,k,i}-\boldsymbol{h}_{n,k}^{H}\boldsymbol{\Theta}\boldsymbol{u}_{i}\|^{2},\label{3-40}\\
&\min_{\mathcal{M}_{2}}~\sum_{i=1}^{N}\|\boldsymbol{u}_{i}-\boldsymbol{G}\boldsymbol{F}\boldsymbol{w}_{i}\|^{2}+\|\boldsymbol{D}-\boldsymbol{F}\boldsymbol{W}\|^{2},\label{3-41}\\
&\min_{\mathcal{M}_{3}}~\|\boldsymbol{D}-\boldsymbol{F}\boldsymbol{W}\|^{2},\label{3-42}
\end{align}
where $\mathcal{M}_{1}$, $\mathcal{M}_{2}$, and $\mathcal{M}_{3}$ are the manifold space defined in the constant modulus constraints in (\ref{3-12e})-(\ref{3-12f}). Then, $\mathcal{M}_{1}$, $\mathcal{M}_{2}$, and $\mathcal{M}_{3}$ are expressed as
\begin{align}
&\mathcal{M}_{1}=\left\{\boldsymbol{\theta}\in\mathbb{C}^{N_{r}\times 1}||\theta_{1}|=\cdots=|\theta_{N_{r}}|=1\right\},\label{3-43}\\
&\mathcal{M}_{2}=\left\{\boldsymbol{F}\in\mathbb{C}^{N_{t}\times N_{RF}}||F_{1,1}|=\cdots=|F_{N_{t},N_{RF}}|=1\right\},\label{3-44}\\
&\mathcal{M}_{3}=\left\{\boldsymbol{D}\in\mathbb{C}^{N_{t}\times N}|\boldsymbol{D}\boldsymbol{D}^{H}\odot\boldsymbol{I}=\boldsymbol{I}\right\},\label{3-45}
\end{align}
{}{In particular,} $\mathcal{M}_{1}$ and $\mathcal{M}_{2}$ are called Riemannian {}{manifolds}, and $\mathcal{M}_{3}$ is {}{called} Oblique manifold\cite{absil2009optimization}. The  principle  of manifold  optimization  method  is  to {}{apply}  the  gradient  descent  algorithm  in {}{the} manifold  space. {}{In particular, the} gradient descent algorithm on Riemannian {}{manifolds} is similar to that in Euclidean {}{spaces}. {}{However, the Riemannian gradient is used for the search direction.} The Riemannian gradients of (\ref{3-40})-(\ref{3-42}) at the current point $\boldsymbol{\theta}$, $\boldsymbol{F}$, $\boldsymbol{D}$ are defined as the projection of {}{the} search direction in {}{the} Euclidean space onto the tangent spaces $\mathcal{T}_{\boldsymbol{\theta}}\mathcal{M}_{1}$, $\mathcal{T}_{\boldsymbol{F}}\mathcal{M}_{2}$, and $\mathcal{T}_{\boldsymbol{D}}\mathcal{M}_{3}$, which can be expressed as  
\begin{align}
&\mathcal{T}_{\boldsymbol{\theta}}\mathcal{M}_{1}=\left\{\boldsymbol{u}\in\mathbb{C}^{N_{r}\times 1}|\mathrm{Re}\left\{\boldsymbol{\theta}^{H}\odot\boldsymbol{u}\right\}=0\right\}\label{3-46},\\
&\mathcal{T}_{\boldsymbol{F}}\mathcal{M}_{2}=\left\{\boldsymbol{U}\in\mathbb{C}^{N\times N_{RF}}|\mathrm{Re}\left\{\boldsymbol{F}^{H}\odot\boldsymbol{U}\right\}=0\right\}\label{3-47},\\
&\mathcal{T}_{\boldsymbol{D}}\mathcal{M}_{3}=\left\{\boldsymbol{V}\in\mathbb{C}^{N_{r}\times N_{RF}}|\mathrm{Re}\left\{\boldsymbol{I}\odot\mathrm{Re}(\boldsymbol{D}\boldsymbol{V}^{H})\right\}=0\right\}\label{3-48}.
\end{align}
where $\odot$ denotes the Hadamard product.
Then, the Euclidean gradients of (\ref{3-40})-(\ref{3-42}) at $\boldsymbol{\theta}$, $\boldsymbol{F}$, and $\boldsymbol{D}$ are computed as follows
\begin{align}
&\nabla_{\boldsymbol{\theta}}f_{1}(\boldsymbol{\theta})=\sum_{n=1}^{N}\sum_{k=1}^{|\mathcal{G}_{n}|}\sum_{i=1}^{N}\mathrm{diag}(\boldsymbol{h}_{n,k}^{H})\boldsymbol{u}_{i}\boldsymbol{u}_{i}\mathrm{diag}(\boldsymbol{h}_{n,k})\boldsymbol{\theta}^{*}
-\boldsymbol{h}_{n,k}^{*}\odot\boldsymbol{u}_{i}v_{n,k,i}^{*},\label{3-49}\\
&\nabla_{\boldsymbol{F}}f_{2}(\boldsymbol{F})=\sum_{i=1}^{N}\boldsymbol{G}^{T}\boldsymbol{G}^{*}\boldsymbol{F}^{*}\boldsymbol{w}_{i}^{*}\boldsymbol{w}_{i}^{T}-\sum_{i=1}^{N}\boldsymbol{G}^{T}\boldsymbol{u}_{i}^{*}\boldsymbol{w}_{i}^{T}+\boldsymbol{F}^{*}\boldsymbol{W}^{*}\boldsymbol{W}^{T}-\boldsymbol{D}^{*}\boldsymbol{W}^{T},\label{3-50}\\
&\nabla_{\boldsymbol{D}}f_{3}(\boldsymbol{D})=\boldsymbol{D}^{*}-\boldsymbol{F}^{*}\boldsymbol{W}^{*},\label{3-51}
\end{align}
where
$f_{1}(\boldsymbol{\theta})=\sum_{n=1}^{N}\sum_{k=1}^{|\mathcal{G}_{n}|}\sum_{k=1}^{N}\|v_{n,k,i}-\boldsymbol{h}_{n,k}^{H}\boldsymbol{\Theta}\boldsymbol{u}_{i}\|$, $f_{2}(\boldsymbol{F})=\sum_{i=1}^{N}\|\boldsymbol{u}_{i}-\boldsymbol{G}\boldsymbol{F}\boldsymbol{w}_{i}\|+\|\boldsymbol{D}-\boldsymbol{F}\boldsymbol{W}\|
$, $f_{3}(\boldsymbol{D})=\|\boldsymbol{D}-\boldsymbol{F}\boldsymbol{W}\|$. Based on the Euclidean gradient, the Riemannian gradients of (\ref{3-40})-(\ref{3-42}) are expressed as
\begin{align}
&\mathrm{grad}_{\boldsymbol{\theta}}~f_{1}(\boldsymbol{\theta})=\nabla_{\boldsymbol{\theta}}f_{1}(\boldsymbol{\theta})-\mathrm{Re}\{\nabla_{\boldsymbol{\theta}}f_{1}(\boldsymbol{\theta})\odot\boldsymbol{\theta}\}\odot\boldsymbol{\theta},\label{3-52}\\
&\mathrm{grad}_{\boldsymbol{F}}~f_{2}(\boldsymbol{F})=\nabla_{\boldsymbol{F}}f_{2}(\boldsymbol{F})-\mathrm{Re}\{\nabla_{\boldsymbol{F}}f_{2}(\boldsymbol{F})\odot\boldsymbol{F}\}\odot\boldsymbol{F},\label{3-53}\\
&\mathrm{grad}_{\boldsymbol{D}}~f_{3}(\boldsymbol{D})=\nabla_{\boldsymbol{D}} f_{3}(\boldsymbol{D})-(\boldsymbol{I}\odot\mathrm{Re}\{\boldsymbol{I}\odot(\nabla_{\boldsymbol{D}} f_{3}(\boldsymbol{D}))^{H}\})\boldsymbol{D}.\label{3-54}
\end{align}
Hence, the current point $\boldsymbol{\theta}$, $\boldsymbol{F}$, and $\boldsymbol{D}$ in the tangent space $\mathcal{T}_{\boldsymbol{\theta}}\mathcal{M}_{1}$, $\mathcal{T}_{\boldsymbol{\theta}}\mathcal{M}_{2}$, $\mathcal{T}_{\boldsymbol{\theta}}\mathcal{M}_{3}$ are updated as $\boldsymbol{\theta}-\delta_{1}\nabla_{\boldsymbol{\theta}}f_{1}(\boldsymbol{\theta})$, $\boldsymbol{F}-\delta_{2}\nabla_{\boldsymbol{F}}f_{2}(\boldsymbol{F})$, and $\boldsymbol{D}-\delta_{3}\nabla_{\boldsymbol{D}}f_{3}(\boldsymbol{D})$, where $\delta_{1}>0$,~$\delta_{2}>0$, and $\delta_{3}>0$ are the step size. It should be noticed that the update point may leave the manifold space. Thus, a retraction operation is used to {}{ensure that} the point {}{stays} in the manifold. {}{More specifically} the retraction operations are expressed as
\begin{align}
&\mathrm{Ret}(\delta_{1}\nabla_{\boldsymbol{\theta}}f_{1}(\boldsymbol{\theta}))=\frac{\boldsymbol{\theta}-\delta_{1}\nabla_{\boldsymbol{\theta}}f_{1}(\boldsymbol{\theta})}{\left\|\boldsymbol{\theta}-\delta_{1}\nabla_{\boldsymbol{\theta}}f_{1}(\boldsymbol{\theta})\right\|},\label{3-55}\\
&\mathrm{Ret}(\delta_{2}\nabla_{\boldsymbol{F}}f_{2}(\boldsymbol{F}))=\frac{\boldsymbol{F}-\delta_{2}\nabla_{\boldsymbol{F}}f_{2}(\boldsymbol{F})}{\left\|\boldsymbol{F}-\delta_{2}\nabla_{\boldsymbol{F}}f_{2}(\boldsymbol{F})\right\|},\label{3-56}\\
&\mathrm{Ret}(\delta_{3}\nabla_{\boldsymbol{F}}f_{3}(\boldsymbol{D}))=\frac{\boldsymbol{D}-\delta_{3}\nabla_{\boldsymbol{D}}f_{3}(\boldsymbol{D})}{\left\|\boldsymbol{D}-\delta_{3}\nabla_{\boldsymbol{D}}f_{3}(\boldsymbol{D})\right\|}.\label{3-57}
\end{align}
Via  these  operations,  we  can  {}{obtain}  the solution {}{for} $\boldsymbol{\Theta},\boldsymbol{F},\boldsymbol{D}$. The details are summarized in \textbf{Algorithm}~\ref{algo3-2}.
\begin{algorithm}
\caption{Proposed AMO Algorithm for Problem (\ref{3-39})}\label{algo3-2} 
\hspace*{0.02in}{\bf Intialization:} The iteration number
$t=0$,~$t_{1}=0$,~$t_{2}=0$,~$t_{3}=0$, the accuracy $\epsilon_{1}$, $\epsilon_{2}$, and $\epsilon_{3}$.\\
\hspace*{0.02in}{\bf Repeat:}\\
\hspace*{0.02in}{\bf Repeat:}\\
Calculate the Euclidean gradient and the Riemannian gradient based on (\ref{3-49}) and (\ref{3-52})\\
Determine the step size $\delta_{1}^{t_{1}}$ based on \cite{absil2009optimization}, then, perform gradient descent algorithm over the current tangent space using $\boldsymbol{\theta}^{t_{1}}-\delta_{1}\nabla_{\boldsymbol{\theta}^{t_{1}}}f_{1}(\boldsymbol{\theta}^{t_{1}})$ and update $\boldsymbol{\theta}$ based on (\ref{3-55})\\
Set $t_{1}=t_{1}+1$.\\
\hspace*{0.02in}{\bf Until:}
$\|\boldsymbol{\theta}^{t_{1}+1}-\boldsymbol{\theta}^{t_{1}}\|\leq \epsilon_{1}$.\\
\hspace*{0.02in}{\bf Repeat:}\\
Calculate the Euclidean gradient and the Riemannian gradient based on (\ref{3-50}) and (\ref{3-53})\\
Determine the step size $\delta_{2}^{t_{2}}$ based on \cite{absil2009optimization}, then, perform gradient descent algorithm over the current tangent space using $\boldsymbol{F}^{t_{2}}-\delta_{2}\nabla_{\boldsymbol{F}^{t_{2}}}f_{2}(\boldsymbol{F}^{t_{2}})$ and update $\boldsymbol{F}^{t_{2}}$ based on (\ref{3-56})\\
Set $t_{2}=t_{2}+1$.\\
\hspace*{0.02in}{\bf Until:}
$\|f_{2}(\boldsymbol{F}^{t_{2}+1})-f_{2}(\boldsymbol{F}^{t_{2}})\|\leq \epsilon_{2}$.\\
\hspace*{0.02in}{\bf Repeat:}\\
Calculate the Euclidean gradient and the Riemannian gradient based on (\ref{3-51}) and (\ref{3-54})\\
Determine the step size $\delta_{3}^{t_{3}}$ based on \cite{absil2009optimization}, then, perform gradient descent algorithm over the current tangent space using $\boldsymbol{D}^{t_{3}}-\delta_{3}\nabla_{\boldsymbol{D}^{t_{3}}}f_{3}(\boldsymbol{D}^{t_{3}})$ and update $\boldsymbol{D}^{t_{3}}$ based on (\ref{3-57})\\
Set $t_{3}=t_{3}+1$.\\
\hspace*{0.02in}{\bf Until:}
$\|f_{3}(\boldsymbol{D}^{t_{3}+1})-f_{3}(\boldsymbol{D}^{t_{3}})\|\leq \epsilon_{3}$.\\
Set $t=t+1$.\\
\hspace*{0.02in}{\bf Until:}
The stop condition is satisfied .\\
\hspace*{0.02in}{\bf Output:}
$\boldsymbol{\theta}^{*}$,$\boldsymbol{F}^{*}$,$\boldsymbol{D}^{*}$.
\end{algorithm}
In the proposed AMO algorithm, the analog beamforming, {}{the} {}{phase shifts} of {}{the} RIS, and {}{the} hybrid beamforming are optimized via the manifold optimization algorithm. According to \textbf{Theorem}~4.3.1 in \cite{absil2009optimization}, the algorithm that uses the manifold optimization method is guaranteed to converge to the point
where the gradient of the objective function is zero \cite{absil2009optimization}.

\subsection{Digital Beamforming Based on {}{the} SCA Algorithm}
{}{We assume that the} hybrid beamforming matrix, {}{the} power allocation matrix, and the phase shift matrix of the RIS are fixed, and the digital beamforming in problem (\ref{3-38}) {}{is} optimized. To  simplify  the  {}{writing}  of  problem (\ref{3-38}),  we  define  the functions
\begin{align}
&f(\{z_{n,k,n}\})=\sum_{n=1}^{N}\log_{2}(z_{n,k,n}\sum_{j=1}^{k}p_{n,j}+\sigma^{2}), ~g(\{v_{n,k,i}\})=-v_{n,k,i}v_{n,k,i}^{H}.\label{3-60}
\end{align}
Also, (\ref{3-38a}) can be re-written as
\begin{subequations}
\begin{align}
\min_{\boldsymbol{W},\{\boldsymbol{u}_{i}\},\{v_{n,k,i}\},\{z_{n,k,i}\}}&\sum_{n=1}^{N}\sum_{k=1}^{|\mathcal{G}_{n}|}\log_{2}(z_{n,k,n}\sum_{j=1}^{k-1}p_{n,j}+\sigma^{2})-\sum_{n=1}^{N}\sum_{k=1}^{|\mathcal{G}_{n}|}\log_{2}(z_{n,k,n}\sum_{j=1}^{k}p_{n,j}+\sigma^{2})+\nonumber\\
&\lambda\Big(\sum_{i=1}^{N}\|\boldsymbol{u}_{i}-\boldsymbol{G}\boldsymbol{F}\boldsymbol{w}_{i}\|^{2}+\|\boldsymbol{D}-\boldsymbol{F}\boldsymbol{W}\|^{2}+\sum_{n=1}^{N}\sum_{k=1}^{|\mathcal{G}_{n}|}\sum_{i=1}^{N}\|v_{n,k,i}-\boldsymbol{h}_{n,k}^{H}\boldsymbol{\Theta}\boldsymbol{u}_{i}\|^{2}\nonumber\\
&+\sum_{n=1}^{N}\sum_{k=1}^{|\mathcal{G}_{n}|}\sum_{i=1}^{N}(z_{n,k,i}-v_{n,k,i}v_{n,k,i}^{H})\Big),\label{3-58a}\\
\mbox{s.t.}~
&\text{(\ref{3-32b})},\text{(\ref{3-32g})},\text{(\ref{3-37d})}.&\label{3-58b}
\end{align}\label{3-58}%
\end{subequations}
The problem in (\ref{3-58}) is a standard DC programming {}{problem}. However, it is still a non-convex {}{problem}. To deal with the non-convex objective function, we use the SCA method \cite{boyd2004convex} {}{in order} to transform the non-convex part of the objective function into a convex function, and then to iteratively solve the convex approximation problem. In the following, we focus our attention on finding convex bounds {}{for} the concave functions  $f(\{z_{n,k,n}\})$ and  $g(\{v_{n,k,i}\})$. {}{To this end, the} first-order Taylor expansion {}{at} the point $(\{z_{n,k,n}\},\{v_{n,k,i}\})$ {}{can be written as}
\begin{align}
&f(\{z_{n,k,n}\}|\{\tilde{z}_{n,k,n}\})=f(\{\tilde{z}_{n,k,n}\}|\{\tilde{z}_{n,k,n}\})+\sum_{n=1}^{N}\sum_{k=1}^{|\mathcal{G}_{n}|}\frac{\sum_{j=1}^{k}p_{n,j}}{\ln 2}\frac{z_{n,k,n}-\tilde{z}_{n,k,n}}{\tilde{z}_{n,k,n}\sum_{j=1}^{k}p_{n,j}+\sigma^{2}}\label{3-61}
\end{align}
and
\begin{eqnarray}
g(\{v_{n,k,i}\}|\{\tilde{v}_{n,k,i}\})=\tilde{v}_{n,k,i}\tilde{v}_{n,k,i}^{H}-2\mathrm{Re}\{v_{n,k,i}\tilde{v}_{n,k,i}^{H}\}.\label{3-62}
\end{eqnarray}
Therefore, the $t+1$th   iteration   of   the proposed   SCA-based iterative  algorithm is expressed as
\begin{subequations}
\begin{align}
\min_{\boldsymbol{W},\{\boldsymbol{u}_{i}\},\{v_{n,k,i}\},\{z_{n,k,i}\}}&\sum_{n=1}^{N}\sum_{k=1}^{|\mathcal{G}_{n}|}\log_{2}(\tilde{z}_{n,k,n}^{t}\sum_{j=1}^{k-1}p_{n,j}+
\sigma^{2})+\sum_{n=1}^{N}\sum_{k=1}^{|\mathcal{G}_{n}|}\frac{\sum_{j=1}^{k}p_{n,j}(z_{n,k,n}-\tilde{z}_{n,k,n}^{t})}{\ln 2(\tilde{z}_{n,k,n}^{t}\sum_{j=1}^{k}p_{n,j}+\sigma^{2})}\nonumber\\
&-\sum_{n=1}^{N}\sum_{n=1}^{N}\sum_{k=1}^{|\mathcal{G}_{n}|}\log_{2}(z_{n,k,n}\sum_{j=1}^{k}p_{n,j}+
\sigma^{2})+\lambda\Big(\sum_{i=1}^{N}\|\boldsymbol{u}_{i}-\boldsymbol{G}\boldsymbol{F}\boldsymbol{w}_{i}\|^{2}\nonumber\\
&+\sum_{n=1}^{N}\sum_{k=1}^{|\mathcal{G}_{n}|}\sum_{i=1}^{N}\|v_{n,k,i}-\boldsymbol{h}_{n,k}^{H}\boldsymbol{\Theta}\boldsymbol{u}_{i}\|^{2}+\|\boldsymbol{D}-\boldsymbol{F}\boldsymbol{W}\|^{2}\nonumber\\
&+\sum_{n=1}^{N}\sum_{k=1}^{|\mathcal{G}_{n}|}\sum_{i=1}^{N}(z_{n,k,i}-\tilde{v}_{n,k,i}^{t}(\tilde{v}_{n,k,i}^{H})^{t}+2\mathrm{Re}\{v_{n,k,i}(\tilde{v}_{n,k,i}^{H})^{t}\})\Big)\label{3-63a}\\
\mbox{s.t.}~
&\text{(\ref{3-32b})},\text{(\ref{3-32g})},\text{(\ref{3-37d})}.&\label{3-63b}
\end{align}\label{3-63}%
\end{subequations}
Starting  with  a  feasible  point  {}{for}  the  problem  (\ref{3-63}),  the  proposed alternating optimization algorithm is summarized in 
\textbf{Algorithm}~\ref{algo3-3}.
\begin{algorithm}
\caption{Proposed SCA-based Algorithm for Problem (\ref{3-63})} \label{algo3-3}
\hspace*{0.02in}{\bf Initialization:}~$t=0$, $\{\tilde{v}_{n,k,j}^{(0)}\}$,$\{\tilde{z}_{n,k,j}^{(0)}\}$, $\epsilon$ is the accuracy.\\
\hspace*{0.02in}{\bf Repeat:}\\
Calculating $\{\boldsymbol{w}_{n}^{(t)}\}$,$\{v_{n,k,j}^{(t)}\}$, $\{z_{n,k,j}^{(t)}\}$,$\{\boldsymbol{u}_{j}^{(t)}\}$ by using CVX \cite{boyd2004convex}.\\
Update $\tilde{v}_{n,k,j}^{(t+1)}=v_{n,k,j}^{(t)}$, $\tilde{z}_{n,k,j}^{(t+1)}=z_{n,k,j}^{(t)}$.\\
Set $t=t+1$;\\
\hspace*{0.02in}{\bf Until:}
$\|\boldsymbol{w}_{n}^{(t)}-\boldsymbol{w}_{n}^{(t)}\|^{2}\leq\epsilon$.\\
\hspace*{0.02in}{\bf Output:}
$\boldsymbol{W}^{*}$\\
\end{algorithm}
{}{The convergence of \textbf{Algorithm}~\ref{algo3-3} is analyzed in the following theorem.}
\begin{theorem}\label{The34}
\textbf{Algorithm}~\ref{algo3-3} converges to a stationary point that satisfies the KKT conditions.

The proof is given in \textbf{Appendix~D}.
\end{theorem} 

\subsection{{}{Optimization Algorithms to Solve (14) and} Computational Complexity}
In the above sections, we have presented the algorithms for power allocation, digital beamforming, analog beamforming, and {}{phase shifts} {}{optimization}. Based on these algorithms. {}{\textbf{Algorithm} \ref{algo3-4}} provides the proposed solution for solving the general optimization problem in (14). In particular, we {}{first} use \textbf{Algorithm} \ref{algo3-1} to solve the power allocation problem, and we assume that the hybrid beamforming matrix and phase shifts of the RIS are fixed.
Thus, the power allocation can be viewed as the function of the power
matrix. Then, we calculate the analog beamforming and {}{phase shifts} of {}{the} RIS by using \textbf{Algorithm} \ref{algo3-2}. Finally, the
digital beamforming matrix {}{is obtained} by using \textbf{Algorithm} \ref{algo3-3}.
\begin{algorithm}[!h]
\caption{Proposed Alternating Optimization Algorithm for Problem (\ref{3-12})} \label{algo3-4}
\hspace*{0.02in}{\bf Initialization:}
$\{p_{m,k}^{(0)}\}$, $\{P_{m}^{(0)}\}$, $\boldsymbol{\Theta}^{(0)}$, $\boldsymbol{W}^{(0)}$, $\boldsymbol{F}^{(0)}$.\\
\hspace*{0.02in}{\bf Repeat:}\\
Using \textbf{Algorithm}\ref{algo3-1} to calculate $\{p_{m,k}^{(t)}\}$ and $\{P_{m}^{(t)}\}$.\\
Using \textbf{Algorithm}\ref{algo3-2} to calculate $\boldsymbol{\Theta}^{(t)}$ and $\boldsymbol{F}^{(t)}$.\\
Using \textbf{Algorithm}\ref{algo3-3} to calculate $\boldsymbol{W}^{(t)}$.\\
Reorder the effective channel gains of the users in each group.\\
Set $t=t+1$.\\
\hspace*{0.02in}{\bf Until:} $p_{m,k}^{(t)}=p_{m,k}^{(t+1)}$, $P_{m}^{(t)}=P_{m}^{(t+1)}$, $\boldsymbol{\Theta}^{(t)}=\boldsymbol{\Theta}^{(t)}$, and $\boldsymbol{W}^{(t)}=\boldsymbol{W}^{(t+1)}$.\\
\hspace*{0.02in}{\bf Output:} 
$\{p_{n,k}^{*}\}$, $\{P_{n}^{*}\}$, $\boldsymbol{\Theta}^{*}$, $\boldsymbol{W}^{*}$, $\boldsymbol{F}^{*}$.
\end{algorithm}

In \textbf{Algorithm} \ref{algo3-1}, the
complexity of calculating the effective channel gains of the users is $\mathcal{O}(MKN)$. {}{Each time that} $\{P_{m}\}$ {}{is updated}, the maximum number of iterations is $M$, and the complexity of computing $\{P_{m}\}$ in each subcycle
is no higher than $\mathcal{O}(K^{2})$. Thus, the complexity of \textbf{Algorithm} \ref{algo3-1} is $\mathcal{O}(MKN+S_{1}MK^{2})$, where $S_{1}$ is the number of iterations. In
\textbf{Algorithm} \ref{algo3-2}, according to \cite{bento2017iteration}, the computational complexity of {}{the} AMO algorithm is {}{$\mathcal{O}(T_{1}\frac{1}{\epsilon_{1}^{2}}+T_{2}\frac{1}{\epsilon_{2}^{2}}+T_{3}\frac{1}{\epsilon_{3}^{2}})$}, where $T_{1}$, $T_{2}$ and $T_{3}$ are the number of iterations.  The transmit beamforming problem {}{in} (\ref{3-63}) is solved by using \textbf{Algorithm}~\ref{algo3-3} {}{that is based on the} SCA method.  Since there are $2N_{t}N_{RF}+2N_{r}K+3NK^{2}$ real variables in problem (\ref{3-63}), the {}{computational} complexity {}{of the} SCA method is $\mathcal{O}(S_{3}(2N_{t}N_{RF}+2N_{r}K+3NK^{2})^{3.5}\log_{2}(\frac{1}{\epsilon}))$ according to \cite{yang2020energy}, where $\epsilon$ is the accuracy of the SCA method and $S_{3}$ is the number of iterations. Therefore, the computational complexity of \textbf{Algorithm}~\ref{algo3-4} is $\mathcal{O}(T(MKN+S_{1}MK^{2}+S_{3}(2N_{t}N_{RF}+2N_{r}K+3NK^{2})^{3.5}\log_{2}(\frac{1}{\epsilon})+T_{1}\frac{1}{\epsilon_{1}^{2}}+T_{2}\frac{1}{\epsilon_{2}^{2}}+T_{3}\frac{1}{\epsilon_{3}^{2}}))$, where $T$ is the number of iterations of \textbf{Algorithm}~\ref{algo3-4}.

\section{Numerical Results}\label{V}

\begin{figure}
\begin{subfigure}{0.5\textwidth}
  \centering
  \includegraphics[height=1.5in,width=2.5in]{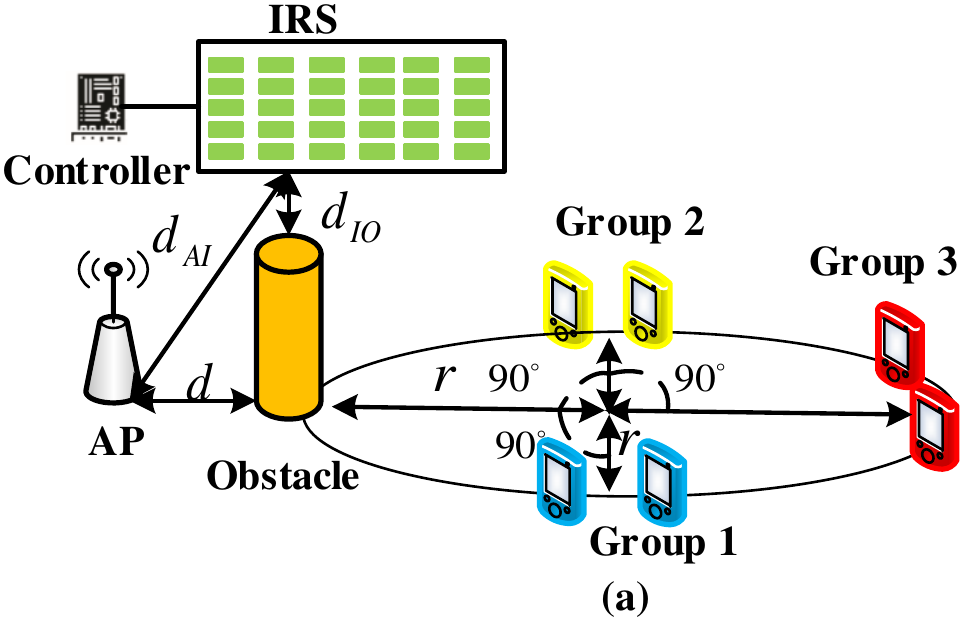}
\end{subfigure}
\begin{subfigure}{0.5\textwidth}
  \centering
  \includegraphics[height=1.5in,width=3.0in]{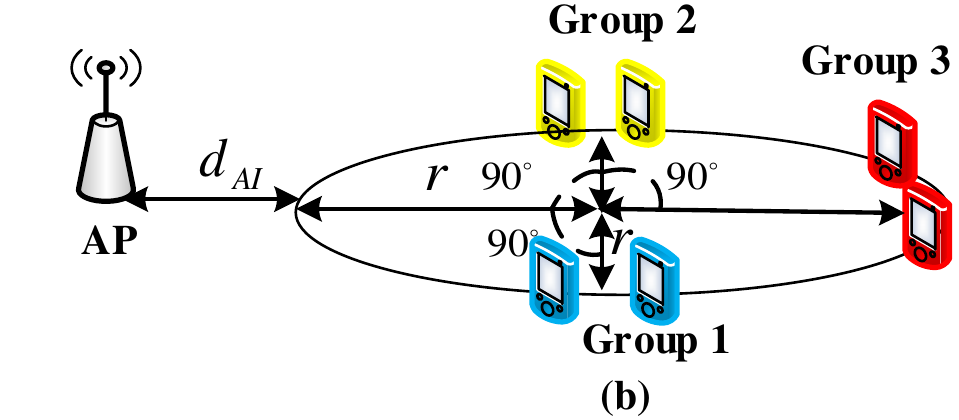}  
\end{subfigure}
\caption{Simulated RIS-aided mmWave-NOMA communication scenario in (a). Simulated mmWave-NOMA communication scenario without RIS in (b).}
\label{fig:3_2}
\end{figure}

In this section, simulation results are provided to verify the performance of the {}{considered} RIS-aided mmWave-NOMA {}{system}. The simulation {}{scenario} is shown in Fig.~\ref{fig:3_2}, {}{where} the obstacles and {}{three groups of users} are distributed on a circle with a radius of $r=50$~m. 
The RIS and the obstacle locate on a {}{line} and the distance between them is $d_{IO} = 9$~m. 
{}{The} AP and {}{the} obstacle are also on the same line and the distance between them is $d=16$~m. 
The distance between {}{the} AP and {}{the} RIS is $25$~m. 
The channel models in (\ref{3-5}) and (\ref{3-6}) are {}{considered}. 
{}{Based on}\cite{akdeniz2014millimeter}, the path fading factor $\alpha$ and $\beta_{n ,k}$ satisfy the Gaussian distribution $\mathcal{CN}(0,10^{-PL_{\alpha}(d_{AI})})$ and $\mathcal{CN}(0,10^{-PL_{\beta_{n,k}}(d_{Ik})})$, where $d_{Ik}$ represents the distance from {}{the} RIS to the $k$ user. 
According to\cite{akdeniz2014millimeter}, $PL_{\alpha}(d_{AI})$ and $PL_{\beta_{n,k}}(d_{Ik})$ can be {}{formulated as follows}
\begin{eqnarray}
PL_{\alpha}(d_{AI})=\eta_{a}+10\eta_{b}\log_{10}(d_{AI})+\beta\\
PL_{\beta_{n,k}}(d_{Ik})=\eta_{a}+10\eta_{b}\log_{10}(d_{Ik})+\beta,
\end{eqnarray}
where $\beta\sim\mathcal{CN}(0,\sigma_{\beta}^{2})$ is the {}{variance of the shadowing}. 
$\eta_{a}=73$, $\eta_{b}=2.92$ and $\sigma_{\beta}=8.7$~dB. 
The other parameters are set as follows: $N_{t}=32$, $N_{RF}=N=3$, $K=6$, $N_{r}=64$. The transmit power {}{is} $P=30$~dBm and the noise power is $\sigma^{2}=-120$~dBm. 
{}{For comparison and benchmarking, four different transmission schemes are considered: a} mmWave-NOMA scheme without RIS, {}{an} RIS-aided all-digital structure mmWave-NOMA system, {}{an} all-digital structure mmWave-NOMA system without RIS, and {}{a} mmWave-FDMA {}{scheme} without RIS. {}{The system setup without RIS is illustrated in} Fig. \ref{fig:3_2}(b). 

\begin{figure}
\begin{subfigure}{0.325\textwidth}
  \centering
  \includegraphics[scale=0.4]{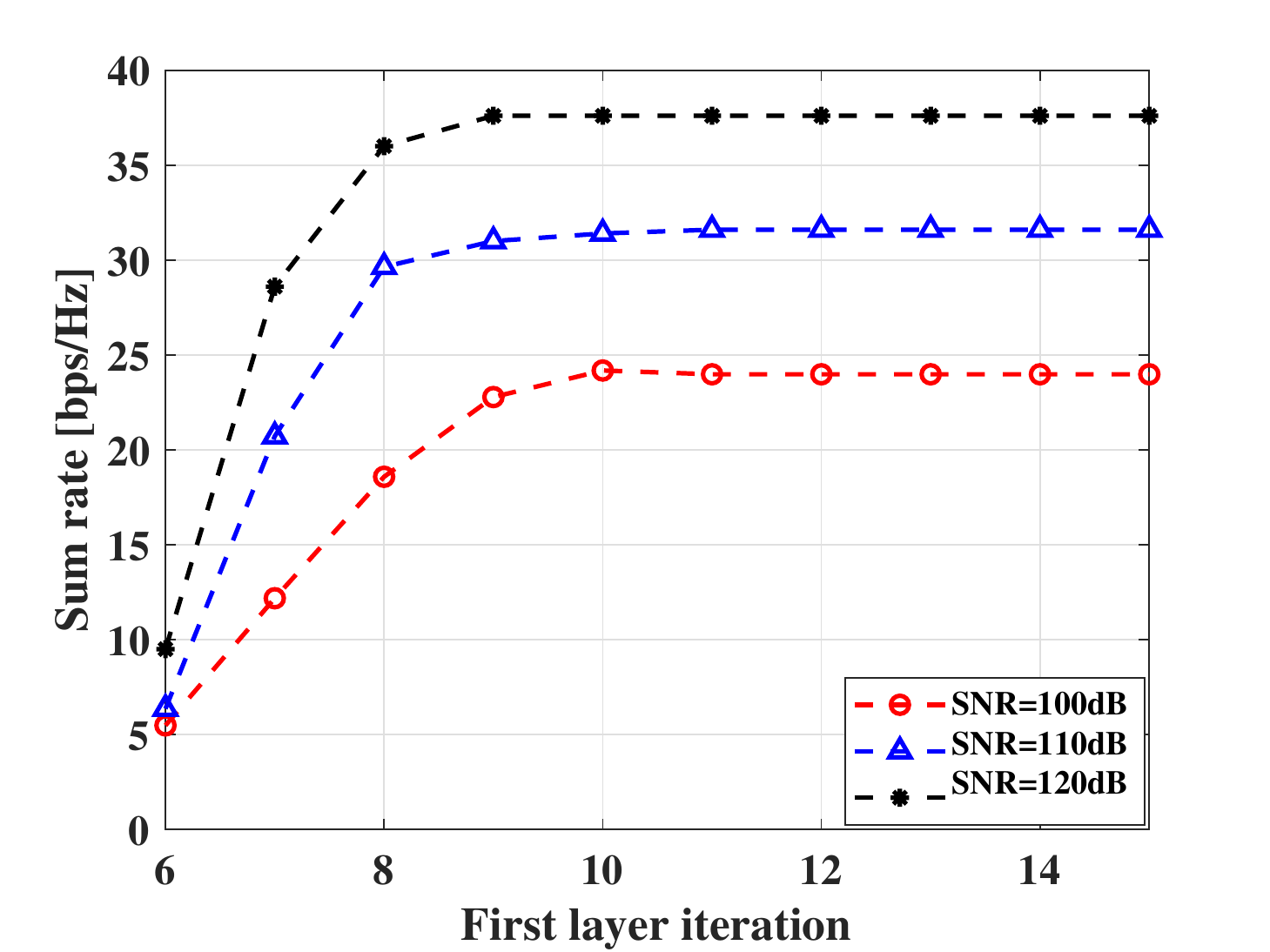}  
  \caption{Convergence behavior of the\\
  proposed \textbf{Algorithm}~\ref{algo3-1},\\ $\gamma=1$~bps/Hz.}
  \label{fig:(3a)}
\end{subfigure}
\begin{subfigure}{0.325\textwidth}
  \centering
  \includegraphics[scale=0.4]{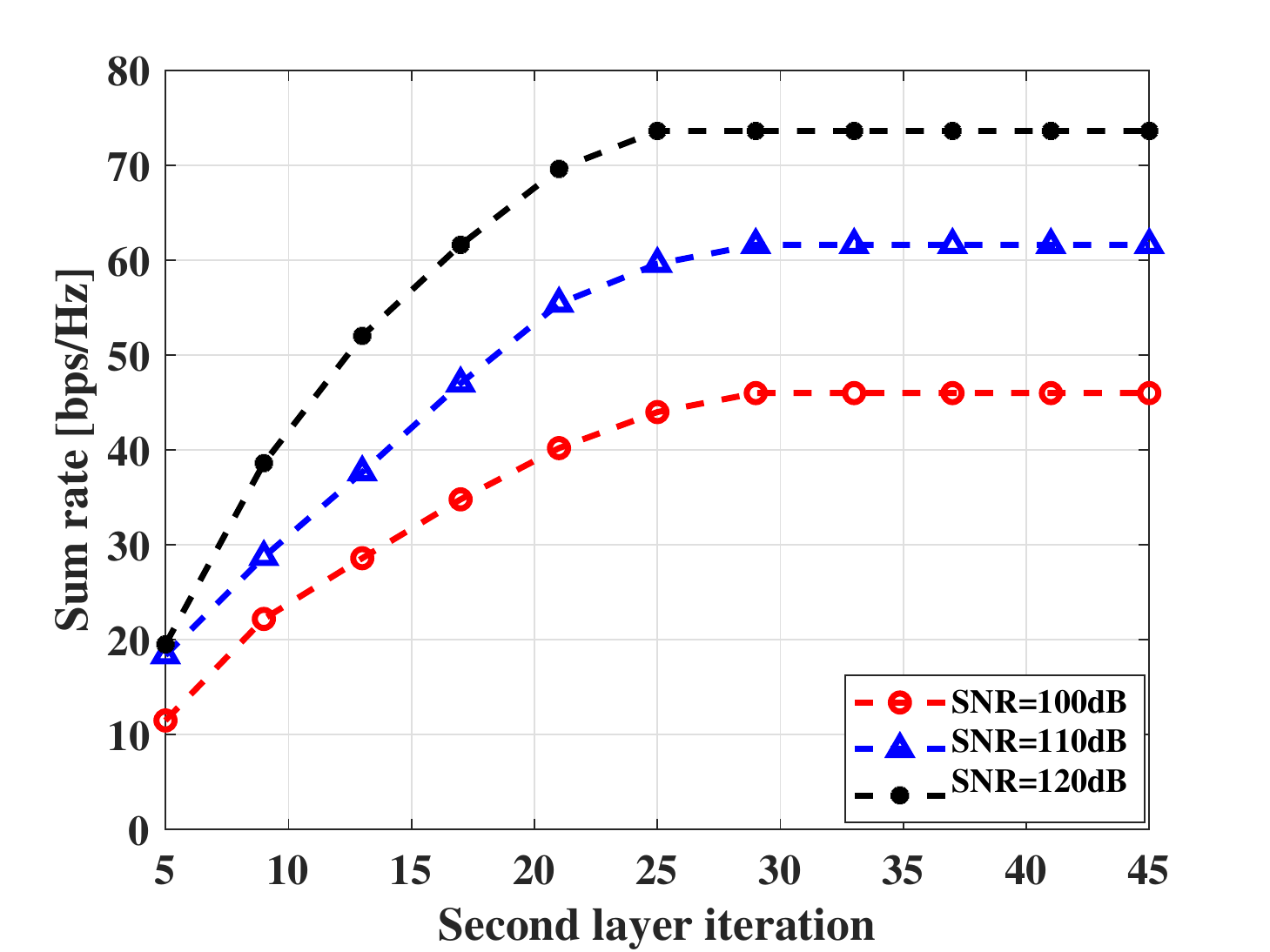}  
  \caption{Convergence behavior of the\\ 
  proposed \textbf{Algorithm}~\ref{algo3-2},\\ $\gamma=1$~bps/Hz.}
  \label{fig:(3b)}
\end{subfigure}
\begin{subfigure}{0.325\textwidth}
  \centering
  \includegraphics[scale=0.4]{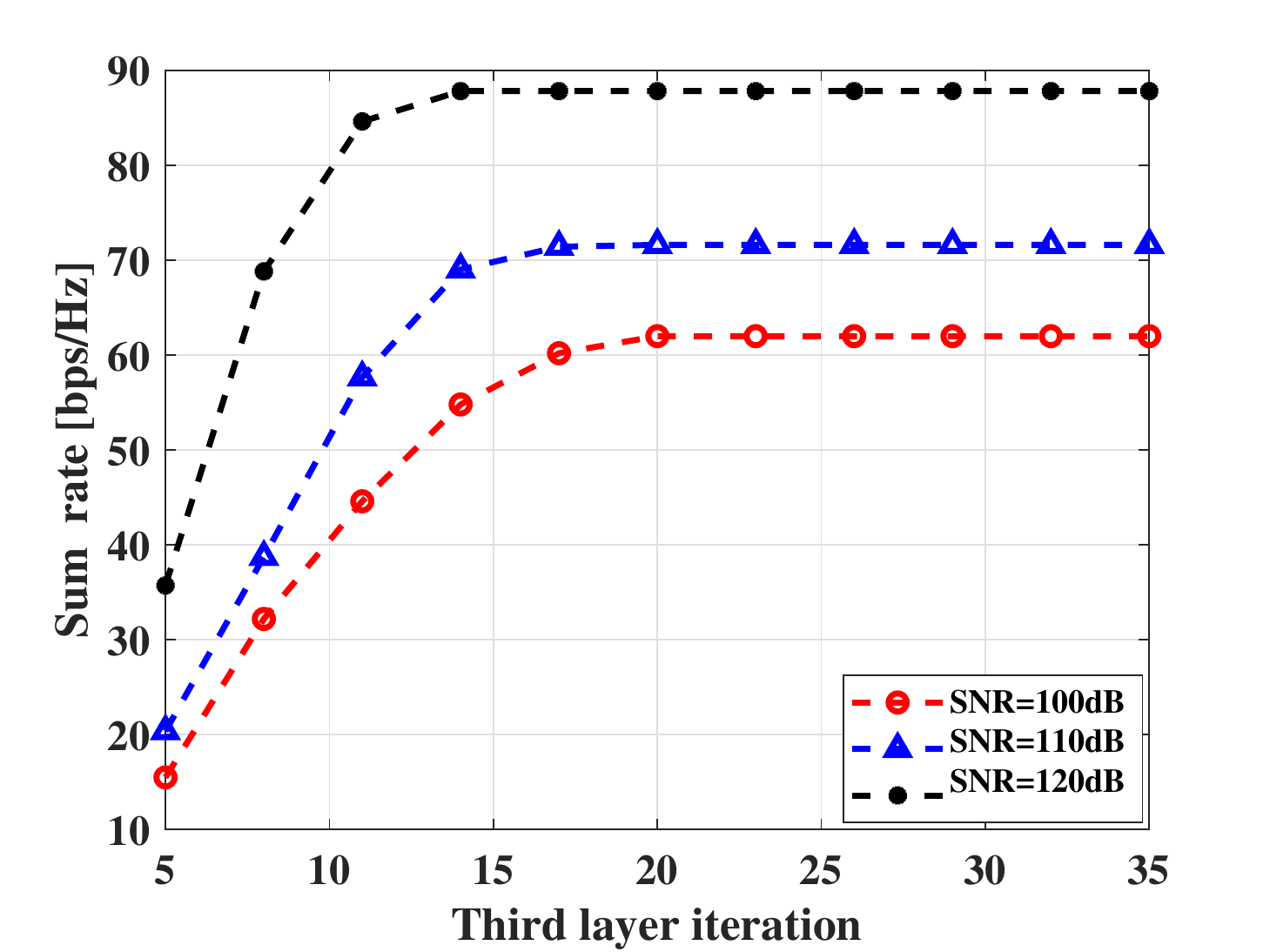}  
  \caption{Convergence behavior of the\\
  proposed \textbf{Algorithm}~\ref{algo3-3},\\ $\gamma=1$~bps/Hz.}
  \label{fig:(3c)}
\end{subfigure}
\caption{Convergence behavior of the proposed algorithms.}
\label{fig:3_6}
\end{figure}

In Figs.~\ref{fig:3_6}, the convergence of the proposed algorithms is analyzed by using numerical {}{simulations}. {}{In the figures, the} power allocation {}{algorithm} is called the first layer iteration, {}{ the} manifold optimization algorithm is called the second layer iteration, and the SCA-based algorithm is called the third iteration. 
In Fig.~\ref{fig:3_6}(a), the convergence of the algorithm against the number of iterations is studied, and it is observed that the sum-rate of the power allocation algorithm converges in about $10$ iterations. The simulation results in {}{Fig. ~\ref{fig:3_6}(b) and Fig. ~\ref{fig:3_6}(c)} show that {}{the phase shifts optimization based on the AMO algorithm} and the transmit beamforming optimization based on SCA algorithm have good convergence performance. The sum-rate increases with the increase of the SNR and finally converges to a stable value. 
\begin{figure}
\begin{subfigure}{.5\textwidth}
  \centering
  \includegraphics[scale=0.45]{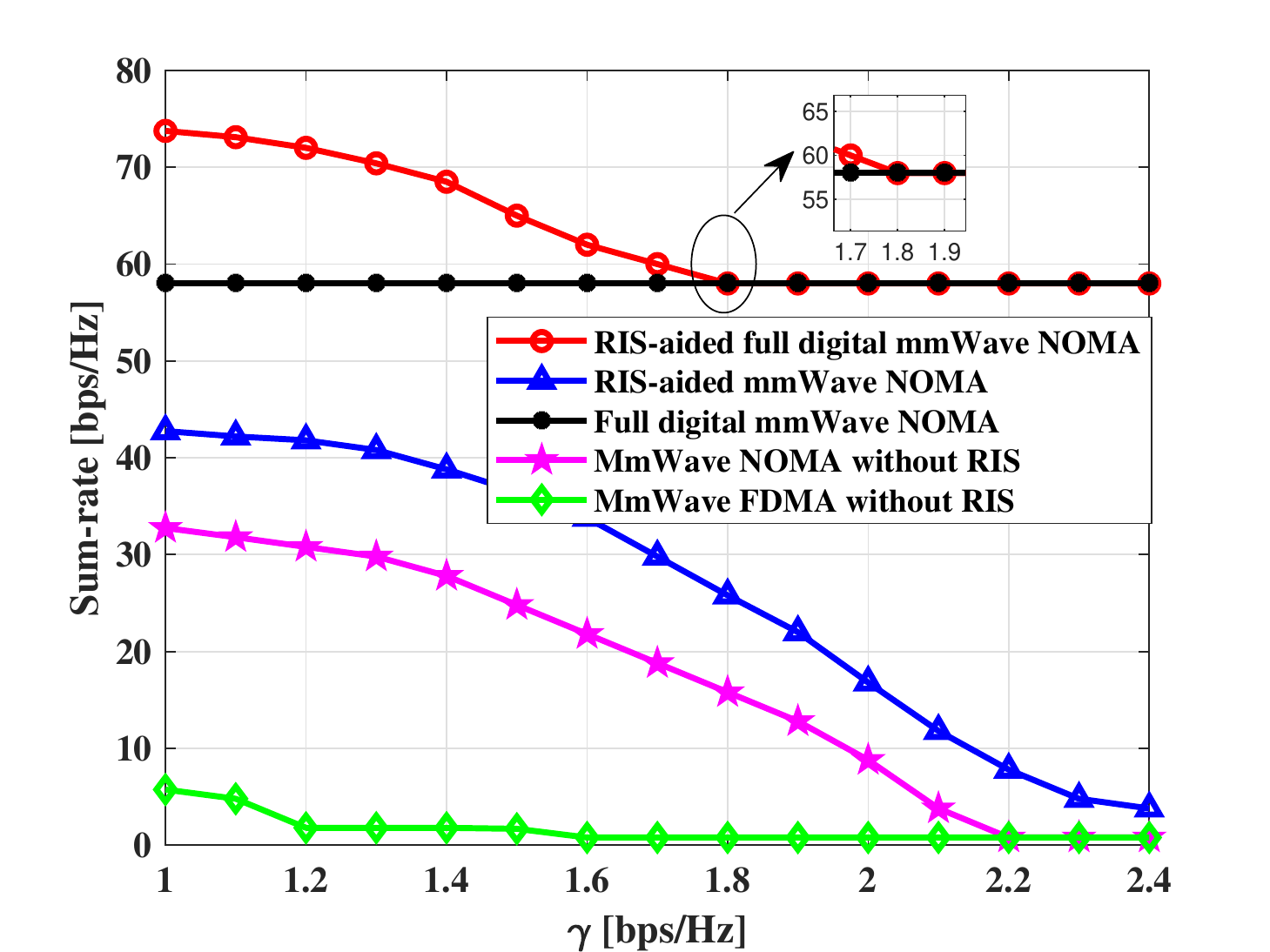}  
  \caption{SNR=100~dB}
  \label{fig:(a)}
\end{subfigure}
\begin{subfigure}{.5\textwidth}
  \centering
  \includegraphics[scale=0.45]{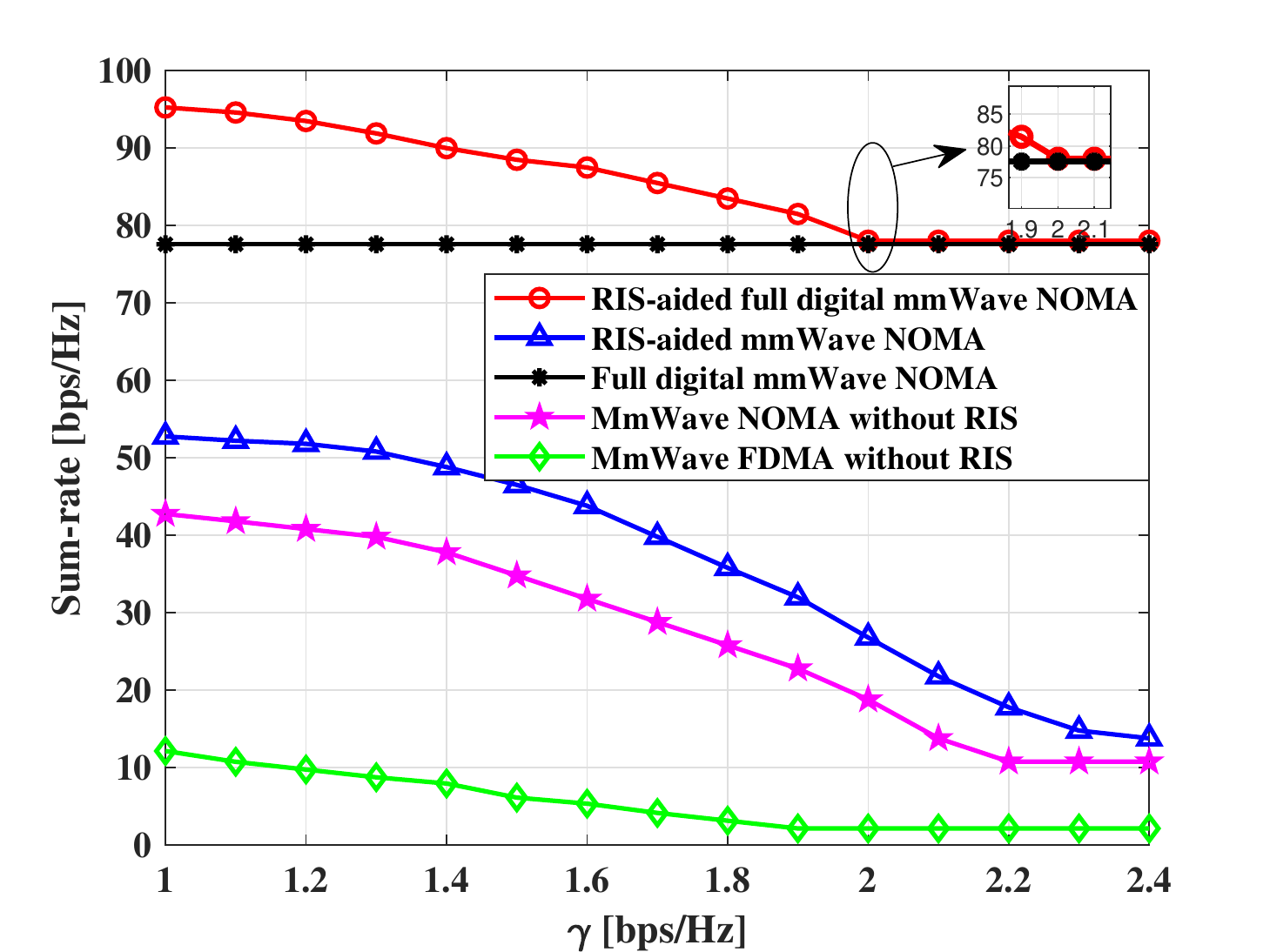}  
  \caption{SNR=120~dB}
  \label{fig:(b)}
\end{subfigure}
\caption{Sum-rate versus minimum rate constraint of user.}
\label{fig:3_7}
\end{figure}


Fig. \ref{fig:3_7} 
{}{compares the sum-rate of all of the considered transmission schemes.} The minimum rate constraint for all the users is equal to $\gamma$. {}{The} proposed RIS-aided hybrid mmWave-NOMA system {}{outperforms} the hybrid mmWave-OMA system without the RIS. Similarly, the RIS-aided fully-digital mmWave-NOMA system {}{outperforms} the fully-digital mmWave-OMA system without the RIS. 
When the minimum rate constraint $\gamma$ is small, e.g., 1 bps, {}{proposed RIS-aided schemes largely outperform the schemes without RIS.}
When $\gamma$ is small, {}{in fact}, more power can be allocated to the users with the highest channel gain in each group by appropriately {}{optimizing the phase shifts of } the RIS. 
Moreover, there {}{exist} channel {}{realizations} that can not satisfy the minimum rate constraint when $\gamma$ is large. In this case, the rate {}{is set equal to zero}. {}{We note} that the sum-rate of the proposed RIS-aided schemes is close to that of all-digital mmWave~NOMA schemes without RIS, which highlights that the proposed RIS-based schemes can suppress the interference well.

\begin{figure}
\begin{subfigure}{.5\textwidth}
  \centering
  \includegraphics[scale=0.45]{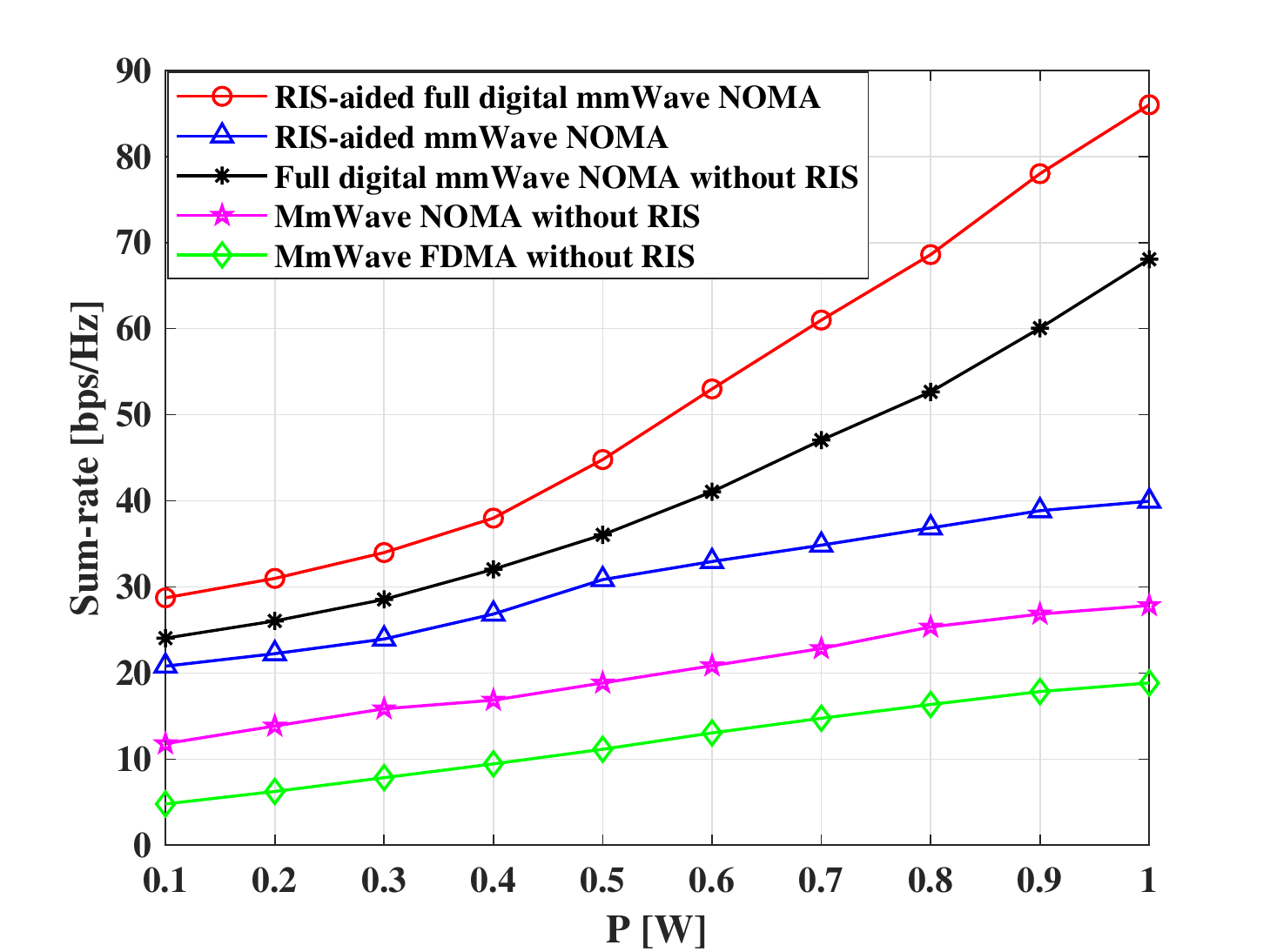}  
  \caption{$\gamma=1$~bps/Hz}
  \label{fig:(3-9a)}
\end{subfigure}
\begin{subfigure}{.5\textwidth}
  \centering
  \includegraphics[scale=0.45]{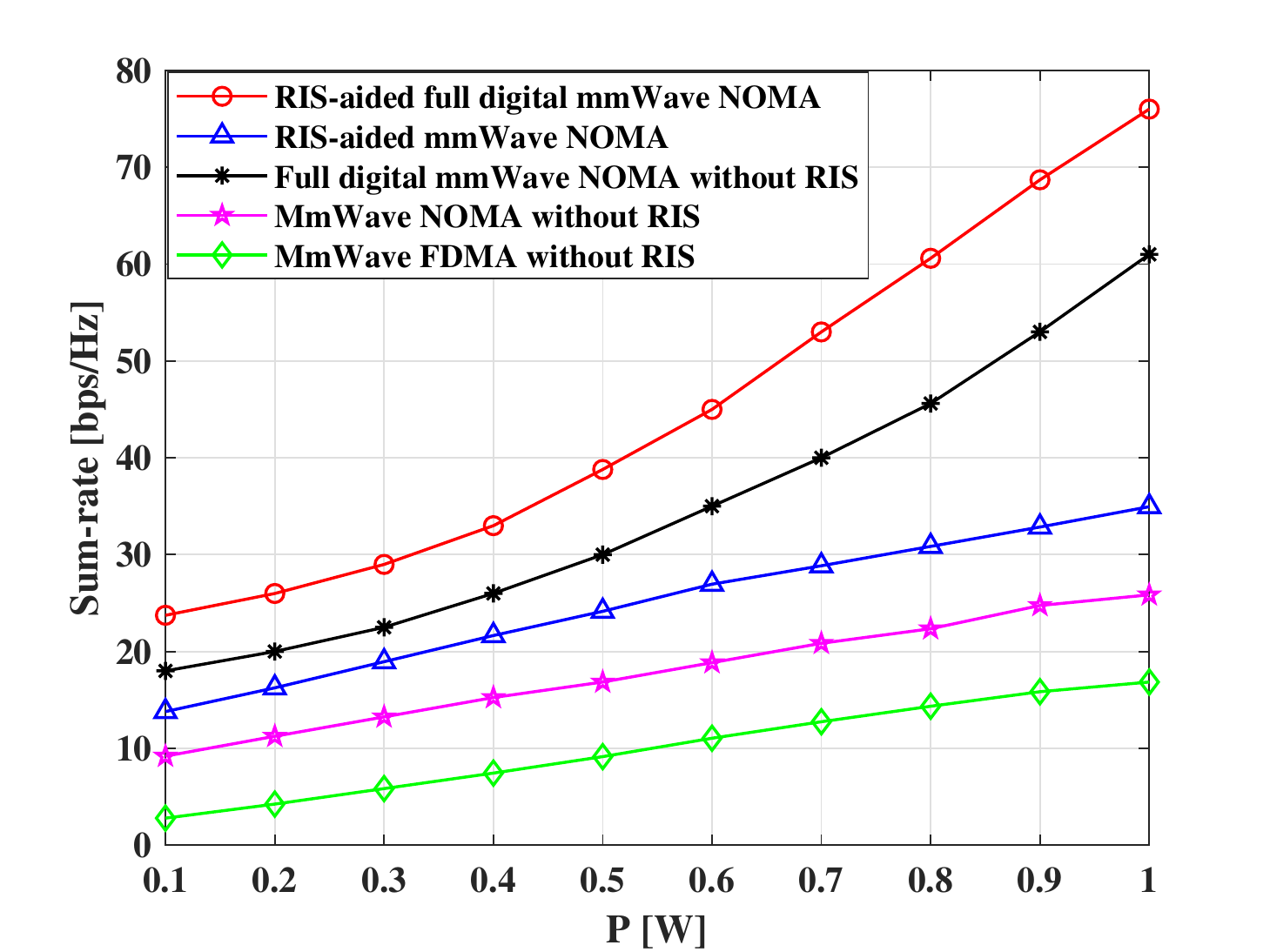}  
  \caption{$\gamma=1.5$~bps/Hz}
  \label{fig:(3-9b)}
\end{subfigure}
\caption{Sum-rate versus total power.}
\label{fig:3_9}
\end{figure}


Fig.~\ref{fig:3_9} 
shows the sum-rate 
of the four {}{considered} schemes {}{as a function of} the total transmission power. 
From Fig.~\ref{fig:3_9}, it can be found that the proposed RIS-aided hybrid mmWave-NOMA and RIS-aided fully-digital mmWave-NOMA {}{schemes} can achieve a higher sum-rate than {}{the} RIS-aided hybrid mmWave-NOMA  and fully-digital mmWave-NOMA systems. In particular, when the transmission power is low, the superiority RIS-aided schemes is more {}{apparent}. 
\begin{figure}
\begin{subfigure}{.5\textwidth}
  \centering
  \includegraphics[scale=0.45]{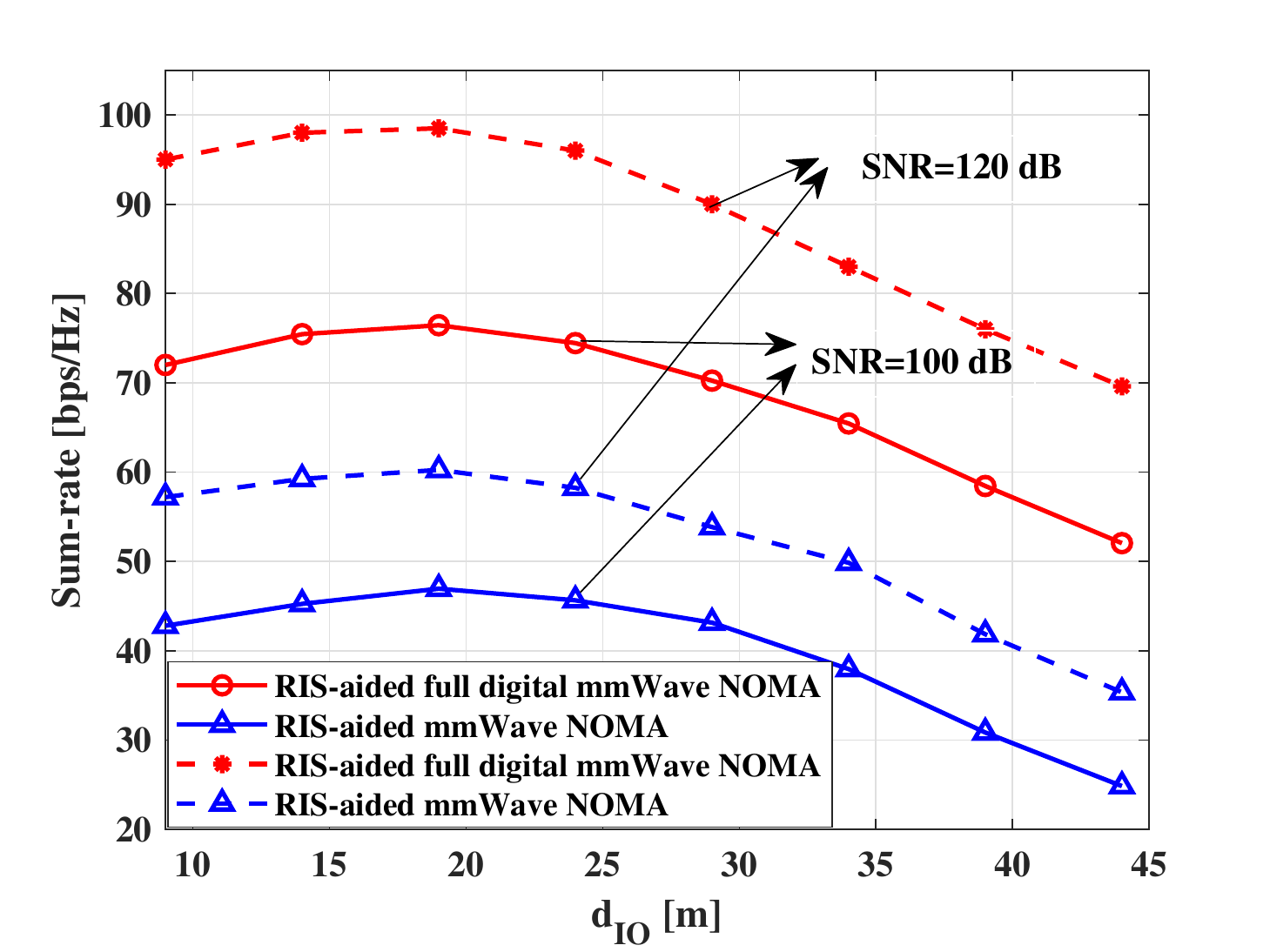}  
  \caption{$P=30$~dBm, $\gamma=1$~bps/Hz}
  \label{fig:(3-10a)}
\end{subfigure}
\begin{subfigure}{.5\textwidth}
\centering
\includegraphics[scale=0.45]{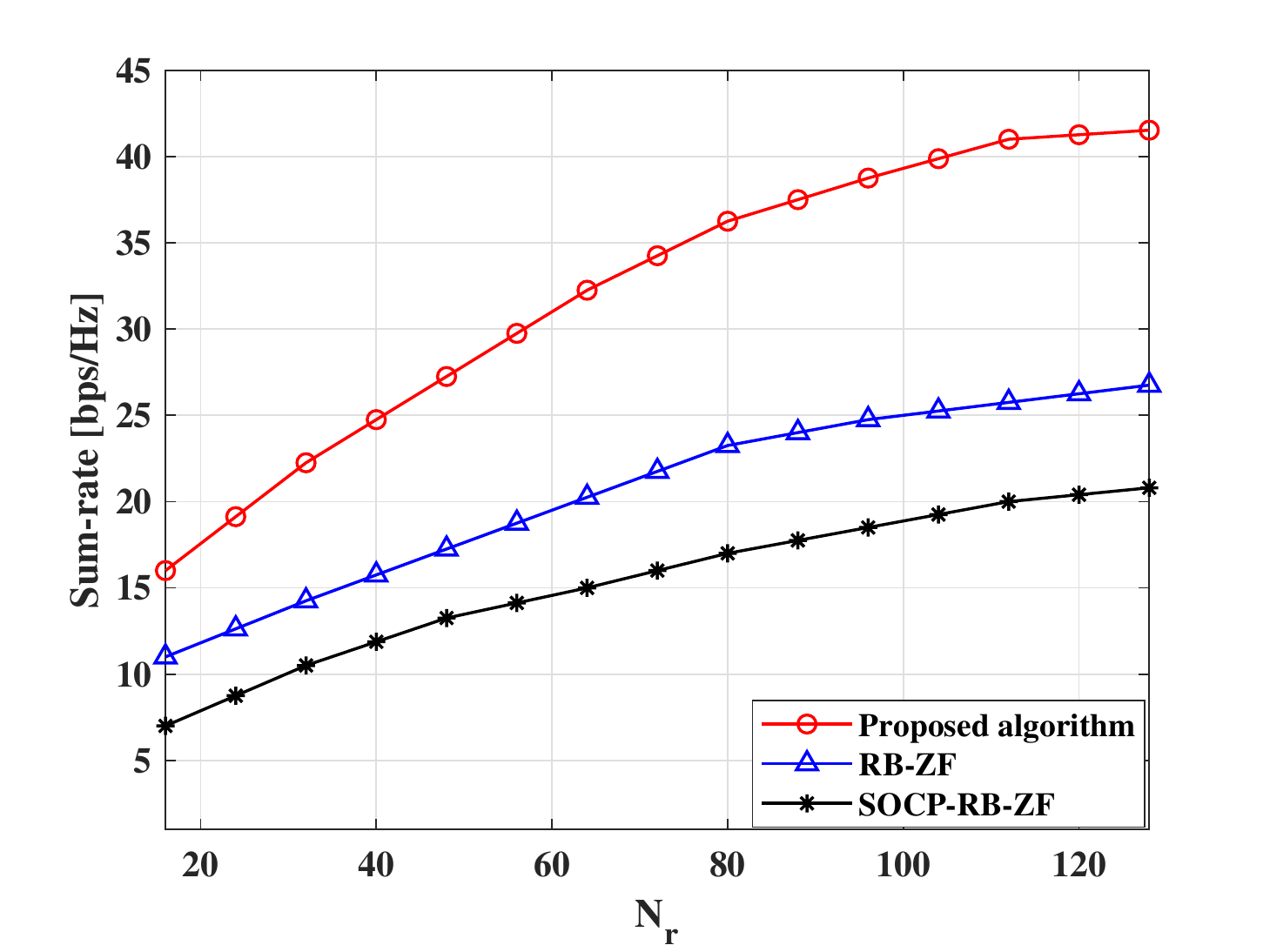}
\caption{$\gamma=1.5$~bps/Hz, $P=0.5$~W.}
\label{fig:3_11}
\end{subfigure}
\caption{Sum-rate of the users versus the distance $d_{IO}$ for
different SNR in (a), sum-rate versus the number of phase shifts of RIS in (b).}
\label{fig:3_10}
\end{figure}

In Fig.~\ref{fig:3_10}(a), we plot the sum-rate of the users against the
distance from the RIS to the obstacle for different SNR, which allows us to examine
the impact of the distance on the system performance.
We compare the performance of the proposed RIS-aided full digital mmWave NOMA
scheme with that of RIS-aided mmWave NOMA scheme. We observe that the sum-rate of the users first
increases with the distance $d_{IO}$ and then decreases
when $d_{IO}$ is beyond a certain value. This is attributed to the fact
that the desired signal and the interference
received by the users decrease when $d_{IO}$ increases. However, the attenuation of the interference is larger than that of desired signal because the interference undergoes a more severe path loss. As $d_{IO}$
further increases, the desired signal strength further
decreases, which decreases the sum-rate of the users. Furthermore,
we observe that the performance gain of all the proposed RIS-aided schemes increase when $d_{IO}$ decreases.
In this situation, the optimization of the location of the RIS becomes critical.

In Fig.~\ref{fig:3_10}(b), the performance of the proposed optimization scheme {}{as a function of} the number of RIS {}{elements is} evaluated. For comparison, the performance of two benchmark schemes is considered {}{as well}. {}{The first benchmark scheme (called RB-ZF in the figure) corresponds to using random phases for the phase shifts of the RIS and for analog beamforming, as well as zero-forcing (ZF) for digital beaforming}, \cite{ali2016non}. The second benchmark scheme corresponds to using (called SOCP-RB-ZF in the figure), analog beamforming with random phase shift and digital precoding used ZF,\cite{li2019joint}. Fig.~\ref{fig:3_11} shows that the proposed RIS-based scheme  has better performance than the other schemes. {}{In all cases, as expected,} the performance improves with the increase {}{of} the number of RIS reflection {}{elements}. 
It is worth noting that the RB-ZF and the SOCP-RB-ZF schemes do not account for the impact of interference from other groups. This is one of the reasons of the superiority of the proposed scheme.

\section{Conclusion}\label{VI}
In this paper, the joint power allocation, phase shifts optimization, and hybrid beamforming design for downlink multiuser RIS-aided mmWave-NOMA was investigated. The phase shifts of the RIS, the power allocation at the AP, and the hybrid beamforming were jointly optimized for maximizing the system sum-rate.  To solve the corresponding non-convex problem, we proposed an alternating optimization algorithm.  First, a sub-optimal algorithm is proposed for power allocation under arbitrarily fixed phase shifts and hybrid beamforming. Then, given the power allocation, we utilized the AMO algorithm and SCA-based algorithm to design the phase shifts of the RIS and the hybrid beamforming to maximize the sum-rate. Finally, numerical results showed that the proposed algorithms are capable of achieving near-optimal performance and that a significant performance gain can be achieved by optimizing the phase shifts of the RIS. In addition,  simulation results showed that the proposed RIS-aided mmWave-NOMA scheme outperforms mmWave-NOMA schemes without the RIS.

\appendices
\section{Proof of \textbf{Theorem~1}}\label{appA}
\setcounter{equation}{64}
Because the interference from other groups
is {}{ignored}, the SINR for user $1$ in group $n$ can be written as follows
\begin{eqnarray}
 f_{n,1}=\frac{|\boldsymbol{h}_{n,1}^{H}\boldsymbol{\Theta}\boldsymbol{G}\boldsymbol{F}\boldsymbol{w}_{n}|^{2}p_{n,1}}{\sigma^{2}}.\label{3-20}
\end{eqnarray}
 Based on $\sum_{k=1}^{|\mathcal{G}_{n}|}p_{n,k}=P_{n}$, the relationship between $p_{n,1}$ and $P_{n}$ is linear. Therefore, the relationship between $\gamma_{n,1}$ and $P_{n}$ is
\begin{eqnarray}
 f_{n,1}=\beta_{n}P_{n}+\alpha_{n},\label{3-21}
\end{eqnarray}
 where $\beta_{n}$ and $\alpha_{n}$ are expressed as
\begin{align}
\beta_{n}=\frac{|\boldsymbol{h}_{n,1}^{H}\boldsymbol{\Theta}\boldsymbol{G}\boldsymbol{F}\boldsymbol{w}_{n}|^{2}}{\sigma^{2}}\left(1-\sum_{k=2}^{|\mathcal{G}_{n}|}\left[(2^{\gamma_{n,k}}-1)\prod_{j=2}^{k}\frac{1}{2^{\gamma_{n,j}}}\right]\right),\label{3-22}
\end{align}
\begin{align}
\alpha_{n}=-\frac{|\boldsymbol{h}_{n,1}^{H}\boldsymbol{\Theta}\boldsymbol{G}\boldsymbol{F}\boldsymbol{w}_{n}|^{2}}{\sigma^{2}}\sum_{k=2}^{|\mathcal{G}_{n}|}\left[(2^{\gamma_{n,k}}-1)\frac{\sigma^{2}}{|\boldsymbol{h}_{n,k}^{H}\boldsymbol{\Theta}\boldsymbol{G}\boldsymbol{F}\boldsymbol{w}_{k}|^{2}}\prod_{j=2}^{k}\frac{1}{2^{\gamma_{n,j}}}\right].\label{3-23}
\end{align}
It is not difficult to find that $\beta_{n}\geq 0$ and $\alpha_{n}\geq 0$. Then, (20a) and (14b) are rewritten as
\begin{eqnarray}
\sum_{n=1}^{N}\log_{2}(\beta_{n}P_{n}+\alpha_{n}+1),
~\eta_{n,1}-\alpha_{n}-P_{n}\beta_{n}\leq 0.\label{3-25}
\end{eqnarray}
According to (\ref{3-25}), the problem in  (20) can be stated as
\begin{subequations}
\begin{align}
\max_{\{P_{n}\}}&\sum_{n=1}^{N}R_{n,1},\label{3-19a1}\\
\mbox{s.t.}~
&\eta_{n,1}-\alpha_{n}-P_{n}\beta_{n}\leq 0, ~\sum_{n=1}^{N}P_{n}\leq 0,&\label{3-19b2}
\end{align}\label{3-191}%
\end{subequations}
We {}{denote} the objective function in (20a) as $f(\{P_{n}\})=\sum_{n=1}^{N}\log_{2}(\beta_{n}P_{n}+\alpha_{n})$. {}{We observe that} $f(\{P_{n}\})$ is {}{an increasing function in} $P_{n}$. {}{Ignoring} the constraint (14b), problem (20) can be solved by the Karush-Kuhn-Tucker (KKT) conditions, i.e.,
\begin{eqnarray}
\frac{\partial f(\{P_{n}\})}{\partial P_{n}}=\lambda,~~\sum_{n=1}^{N}P_{n}=P.\label{3-69}
\end{eqnarray}
The solution of equation in (\ref{3-69}) is
\begin{eqnarray}
\bar{P}_{n}=\frac{P}{N}-\frac{\alpha_{n}+1}{\beta_{n}}+\sum_{i=1}^{N}\frac{\alpha_{i}+1}{N\beta_{i}}.
\end{eqnarray}
Then, \textbf{Theorem~1} is proved.

\section{Proof of \textbf{Theorem~2}}\label{appB}
To prove \textbf{Theorem~2}, we use {}{the method of} proof by contradiction. Assuming there is an optimal solution $\hat{P}_{n1}$ which satisfies $\hat{P}_{n_{1}}>\frac{2^{\gamma_{n,1}}-\alpha_{n}}{\beta_{n}}>\bar{P}_{n_{1}}$. Since $\sum_{n=1}^{N}\hat{P}_{n}\leq P$, there always exists $\hat{P}_{n_{2}}\leq \bar{P}_{n_{2}}$ ($n_{2}\neq n_{1}$). The power allocation solution can be expressed as
\begin{eqnarray}
L_{n_{1}}=\hat{P}_{n_{1}}-\delta,~L_{n_{2}}=\hat{P}_{n_{2}}+\delta, ~L_{n}=\hat{P}_{n}, n\neq n_{1}, n_{2}.
\end{eqnarray}
Therefore, {}{we} only need to prove that the sum rate of the power allocation solution $\{L_{n}\}$ is higher than that of the solution $\{\hat{P}_{n}\}$.  We assume {}{that} the objective function in (20a) with solution $\{L_{n}\}$ {}{is} $g(\{L_{n}\})=\sum_{n=1}^{N}\log_{2}(\beta_{n}L_{n}+\alpha_{n})$, and the objective function in (20a) with solution $\{\hat{P}_{n}\}$ is $h(\{\hat{P}_{n}\})=\sum_{n=1}^{N}\log_{2}(\beta_{n}\hat{P}_{n}+\alpha_{n})$. {}{If} $\delta=0$, then $g(\{L_{n}\})-h(\{\hat{P}_{n}\})=0$. The derivative of $g(\{L_{n}\})-h(\{\hat{P}_{n}\})$ with respect to $\delta$ is 
\begin{align}
\frac{\partial g(\{L_{n}\})-h(\{\hat{P}_{n}\})}{\partial \delta}=
\frac{1}{\ln 2}\frac{\beta_{n_{2}}}{(\beta_{n_{2}}(\hat{P}_{n_{2}}+\delta)+\alpha_{n_{2}}+1)}-\frac{1}{\ln 2}\frac{\beta_{n_{1}}}{(\beta_{n_{1}}(\hat{P}_{n_{1}}-\delta)+\alpha_{n_{1}}+1)}.
\end{align}
According to the derivative of the objective function (20a), we have
\begin{align}
\frac{1}{\ln 2}\frac{\beta_{n_{2}}}{(\beta_{n_{2}}(\hat{P}_{n_{2}}+\delta)+\alpha_{n_{2}}+1)}>\frac{1}{\ln 2}\frac{\beta_{n_{2}}}{(\beta_{n_{2}}\bar{P}_{n_{2}}+\alpha_{n_{2}}+1)}\\
\frac{1}{\ln 2}\frac{\beta_{n_{1}}}{(\beta_{n_{1}}(\hat{P}_{n_{1}}-\delta)+\alpha_{n_{1}}+1)}<\frac{1}{\ln 2}\frac{\beta_{n_{1}}}{(\beta_{n_{1}}\bar{P}_{n_{1}}+\alpha_{n_{1}}+1)},
\end{align}
where $\frac{1}{\ln 2}\frac{\beta_{n_{1}}}{(\beta_{n_{1}}\bar{P}_{n_{1}}+\alpha_{n_{1}}+1)}=\frac{1}{\ln 2}\frac{\beta_{n_{2}}}{(\beta_{n_{2}}\bar{P}_{n_{2}}+\alpha_{n_{2}}+1)}=\frac{1}{\ln 2}\frac{N}{P+\sum_{n=1}^{N}\frac{\alpha_{n}+1}{\beta_{n}}}$.
Therefore, (72) is equivalent to 
\begin{align}
\frac{\partial g(\{L_{n}\})-h(\{\hat{P}_{n}\})}{\partial \delta}>\frac{1}{\ln 2}\frac{\beta_{n_{2}}}{(\beta_{n_{2}}\bar{P}_{n_{2}}+\alpha_{n_{2}}+1)}-\frac{1}{\ln 2}\frac{\beta_{n_{1}}}{(\beta_{n_{1}}\bar{P}_{n_{1}}+\alpha_{n_{1}}+1)}=0.
\end{align}
Since $g(\{L_{n}\})-h(\{\hat{P}_{n}\})=0$, we have
$g(\{L_{n}\})>h(\{\hat{P}_{n}\})$, which demonstrates that $\{L_{n}\}$ is better than $\{\hat{P}_{n}\}$. {}{This} contradicts the assumption that $\{\hat{P}_{n}\}$ is the optimal solution. To this end, the optimal solution of problem (20) must satisfy $\hat{P}_{n}=\frac{2^{\gamma_{n,1}-\alpha_{n}}}{\beta_{n}},~n\in\mathcal{N}$. {}{Thus,} \textbf{Theorem~2} is proved.



\section{Proof of \textbf{Theorem 3}}\label{appC}
According to {}{the} Schur complement, $\left[
\begin{matrix}
  z_{n,k,i} & v_{n,k,i} \\
   v_{n,k,i}^{H} & 1\\
   \end{matrix}
  \right]\succeq\boldsymbol{0}$ is equivalent to 
$z_{n,k,i}-v_{n,k,i}v_{n,k,i}^{H}>0$.
Combining $z_{n,k,i}-v_{n,k,i}v_{n,k,i}^{H}>0$ with $z_{n,k,i}-v_{n,k,i}v_{n,k,i}^{H}<0$, we have $z_{n,k,i}-v_{n,k,i}v_{n,k,i}^{H}=0$, thus, $z_{n,k,i}=v_{n,k,i}v_{n,k,i}^{H}$, which demonstrates that we can replace (30f) with (33) and (34). {}{Thus} \textbf{Theorem~3} is proved.

\section{Proof of \textbf{Theorem~4}}\label{appD}
In order to simplify the notation, let  
$Q(\{\boldsymbol{w}_{i}\},\{\boldsymbol{u}_{i}\},\{v_{n,k,i}\},\{z_{n,k,i}\})$ denote (57a)
and
$Q(\{\boldsymbol{w}_{i}\},$ $\{\boldsymbol{u}_{i}\},\{v_{n,k,i}\},\{z_{n,k,i}\}|\{\tilde{v}_{n,k,i}\},\{\tilde{z}_{n,k,i}\}|)$ denote (60a).
In each iteration of \textbf{Algorithm 4}, $Q(\{\boldsymbol{w}_{i}\},$ $\{\boldsymbol{u}_{i}\},\{v_{n,k,i}\},\{z_{n,k,i}\})$ is replaced by $Q(\{\boldsymbol{w}_{i}\},\{\boldsymbol{u}_{i}\},$ $\{v_{n,k,i}\},\{z_{n,k,i}\}|\{\tilde{v}_{n,k,i}\},\{\tilde{z}_{n,k,i}\}|)$, which is a differentiable convex function. Based on \cite{marks1978general}, \textbf{Algorithm 4} converges to a KKT point of the problem (58), which must satisfy the following conditions:
\begin{eqnarray}
&Q(\{\boldsymbol{w}_{i}\},\{\boldsymbol{u}_{i}\},\{v_{n,k,i}\},\{z_{n,k,i}\})\leq Q(\{\boldsymbol{w}_{i}\},\{\boldsymbol{u}_{i}\},\{v_{n,k,i}\},\{z_{n,k,i}\}|\{\tilde{v}_{n,k,i}\},\{\tilde{z}_{n,k,i}\}|)\label{3-79}\\
&Q(\{\boldsymbol{w}_{i}^{t}\},\{\boldsymbol{u}_{i}^{t}\},\{v_{n,k,i}^{t}\},\{z_{n,k,i}^{t}\})=Q(\{\boldsymbol{w}_{i}^{t}\},\{\boldsymbol{u}_{i}^{t}\},\{v_{n,k,i}^{t}\},\{z_{n,k,i}^{t}\}|\{\tilde{v}_{n,k,i}^{t+1}\},\{\tilde{z}_{n,k,i}^{t+1}\}|)\label{3-80}\\
&\frac{\partial Q(\{\boldsymbol{w}_{i}^{t}\},\{\boldsymbol{u}_{i}^{t}\},\{v_{n,k,i}^{t}\},\{z_{n,k,i}\})}{\partial v_{n,k,i}}=\frac{\partial Q(\{\boldsymbol{w}_{i}^{t}\},\{\boldsymbol{u}_{i}^{t}\},\{v_{n,k,i}^{t}\},\{z_{n,k,i}^{t}\}|\{\tilde{v}_{n,k,i}^{t+1}\},\{\tilde{z}_{n,k,i}^{t+1}\}|)}{\partial v_{n,k,i}}\label{3-83}\\
&\frac{\partial Q(\{\boldsymbol{w}_{i}^{t}\},\{\boldsymbol{u}_{i}^{t}\},\{v_{n,k,i}^{t}\},\{z_{n,k,i}^{t}\})}{\partial z_{n,k,i}}=\frac{\partial Q(\{\boldsymbol{w}_{i}^{t}\},\{\boldsymbol{u}_{i}^{t}\},\{v_{n,k,i}^{t}\},\{z_{n,k,i}^{t}\}|\{\tilde{v}_{n,k,i}^{t+1}\},\{\tilde{z}_{n,k,i}^{t+1}\}|)}{\partial z_{n,k,i}}\label{3-84}.
\end{eqnarray}
According to the {}{Taylor} expansion, $Q(\{\boldsymbol{w}_{i}\},\{\boldsymbol{u}_{i}\},\{v_{n,k,i}\},\{z_{n,k,i}\})$ and $Q(\{\boldsymbol{w}_{i}\},\{\boldsymbol{u}_{i}\},\{v_{n,k,i}\},$ $\{z_{n,k,i}\}|\{\tilde{v}_{n,k,i}\},\{\tilde{z}_{n,k,i}\}|)$ satisfy the first condition.
{}{Since} $\tilde{v}_{n,k,i}^{t+1}=v_{n,k,i}^{t}$ and $\tilde{z}_{n,k,i}^{t+1}=z_{n,k,i}^{t}$, the second condition is satisfied as well.
Finally, we verify the conditions in (\ref{3-83})-(\ref{3-84}) by deriving the first derivatives of $Q(\{\boldsymbol{w}_{i}^{t}\},\{\boldsymbol{u}_{i}^{t}\},\{v_{n,k,i}^{t}\},\{z_{n,k,i}\})$ and $Q(\{\boldsymbol{w}_{i}^{t}\},\{\boldsymbol{u}_{i}^{t}\},\{v_{n,k,i}^{t}\},\{z_{n,k,i}^{t}\}|\{\tilde{v}_{n,k,i}^{t+1}\}$ $,\{\tilde{z}_{n,k,i}^{t+1}\}|)$ with respect to $v_{n,k,i}$, and $z_{n,k,i}$, respectively. {}{They can be written as follows:}
\begin{align}
&\frac{\partial Q(\{\boldsymbol{w}_{i}^{t}\},\{\boldsymbol{u}_{i}^{t}\},\{v_{n,k,i}^{t}\},\{z_{n,k,i}^{t}\})}{\partial v_{n,k,i}}=\lambda((v_{n,k,i}^{t})^{*}-((\boldsymbol{h}_{n,k}^{t})^{H}\boldsymbol{\Theta}^{t}\boldsymbol{u}^{t})^{*}-(v_{n,k,i}^{t})^{*}).
\end{align}
\begin{align}
&\frac{\partial Q(\{\boldsymbol{w}_{i}^{t}\},\{\boldsymbol{u}_{i}^{t}\},\{v_{n,k,i}^{t}\},\{z_{n,k,i}^{t}\}|\{\tilde{v}_{n,k,i}^{t+1}\},\{\tilde{z}_{n,k,i}^{t+1}\}|)}{\partial v_{n,k,i}}=\lambda((v_{n,k,i}^{t})^{*}-((\boldsymbol{h}_{n,k}^{t})^{H}\boldsymbol{\Theta}^{t}\boldsymbol{u}^{t})^{*}-(v_{n,k,i}^{t})^{*}).
\end{align}
\begin{align}
&\frac{\partial Q(\{\boldsymbol{w}_{i}^{t}\},\{\boldsymbol{u}_{i}^{t}\},\{v_{n,k,i}^{t}\},\{z_{n,k,i}^{t}\})}{\partial z_{n,k,i}}=\nonumber\\
&\begin{cases}
\frac{\sum_{j=1}^{k-1}p_{n,j}}{\ln_{2}(z_{n,k,n}^{t}\sum_{j=1}^{k-1}p_{n,j}+\sigma^{2})}-\frac{\sum_{j=1}^{k}p_{n,j}}{\ln_{2}(z_{n,k,n}^{t}\sum_{j=1}^{k}p_{n,j}+\sigma^{2})}& n=i\\
1& n\neq i.
\end{cases}
\end{align}
\begin{align}
&\frac{\partial Q(\{\boldsymbol{w}_{i}^{t}\},\{\boldsymbol{u}_{i}^{t}\},\{v_{n,k,i}^{t}\},\{z_{n,k,i}^{t}\}|\{\tilde{v}_{n,k,i}^{t+1}\},\{\tilde{z}_{n,k,i}^{t+1}\}|)}{\partial z_{n,k,i}}=\nonumber\\
&\begin{cases}
\frac{\sum_{j=1}^{k-1}p_{n,j}}{\ln_{2}(z_{n,k,n}^{t}\sum_{j=1}^{k-1}p_{n,j}+\sigma^{2})}-\frac{\sum_{j=1}^{k}p_{n,j}}{\ln_{2}(z_{n,k,n}^{t}\sum_{j=1}^{k}p_{n,j}+\sigma^{2})}& n=i\\
1& n\neq i.
\end{cases}
\end{align}
The conditions (20) and (21) are verified to be satisfied. Therefore, $\textbf{Algorithm 4}$ converges to a KKT solution of problem (58).
\bibliographystyle{IEEEtran}

\bibliography{relate}

\end{document}